\documentclass [a4paper,11pt]{article}
\pdfoutput=1
\usepackage[english]{babel}
\usepackage{graphicx,bm,natbib,amsmath,amsfonts,amssymb,hyperref,slashed,upgreek,caption}
\newcommand{\be}{\begin{equation}}
\newcommand{\ee}{\end{equation}}
\newcommand{\bea}{\begin{eqnarray}}
\newcommand{\eea}{\end{eqnarray}}
\newcommand{\de}{\partial}
\newcommand{\HH}{{\cal H}}

\newcommand{\bt}{\beta}
\newcommand{\gam}{\gamma}

\newcommand{\del}{\delta}

\newcommand{\Om}{\Omega}
\newcommand{\rmd}{\textrm{d}}

\newcommand{\avg}[1]{\langle #1 \rangle}

\def\ie{{\em i.e.}~}

\def\d{\delta}
\def\tt{\Uptheta}
\def\kv{\vec k}
\def\qv{\vec q}

\def\vv{\vec v}

\def\kMpc{\, h \, {\rm Mpc}^{-1}}

\begin{document}
\def\thefootnote{\fnsymbol{footnote}}

\begin{center}
\Large{\textbf{The nonlinear power spectrum\\ in clustering quintessence cosmologies}} \\[0.4cm]
 
\large{Guido D'Amico\footnote{email: \texttt{gda2@nyu.edu}}}
\\[0.4cm]

\small{
\textit{Center for Cosmology and Particle Physics\\ New York University, 4 Washington place, New York, NY 10003, USA}}
\\[0.4cm]

\large{Emiliano Sefusatti\footnote{email: \texttt{emiliano.sefusatti@cea.fr}}}
\\[0.4cm]

\small{
\textit{Institut de Physique Th\'eorique\\ CEA, IPhT, 91191 Gif-sur-Yvette c\'edex, France\\ CNRS, URA-2306, 91191 Gif-sur-Yvette c\'edex, France}}

\end{center}

\vspace{.5cm}

\hrule \vspace{0.3cm}
\noindent \small{\textbf{Abstract}
\vspace{0.2cm}\\
\noindent We study the nonlinear evolution of density perturbations in cosmologies where the late-time accelerated expansion is driven by a quintessence field with vanishing speed of sound. For these models matter and quintessence perturbations are comoving and it is possible to write a single continuity equation for the {\em total} density fluctuations given by a weighted sum of the two components. Including the Euler equation for the common velocity field we solve the evolution equations for the nonlinear, total density power spectrum in the Time-Renormalization Group approach. In fact any cosmological observable is directly related by gravity only to the {\em total} density perturbations, with the two components being individually unobservable. We show that the clustering of quintessence perturbations induces small corrections with respect to the nonlinear evolution of power spectrum in smooth quintessence models described by the same equation of state. Such small corrections, however, contrast with the large effect of a vanishing speed of sound on the linear growth function at low redshift. For this reason, models with the same normalization of the linear density power spectrum can present significantly different nonlinear corrections depending on the value of the sound speed. Although such differences vanish in the $w\rightarrow-1$, $\Lambda$CDM limit, we argue that the relation between linear and nonlinear growth of structures should be properly taken into account in constraining models with inhomogeneous dark energy. \\
\hrule
\def\thefootnote{\arabic{footnote}}
\setcounter{footnote}{0}


\section{Introduction}

The nature of dark energy, \ie the source of the current accelerated expansion of the Universe, is one of the deepest mysteries in contemporary cosmology. Although the simplest explanation is the presence of an (unnaturally) tiny cosmological constant, it is important to explore other scenarios both from the point of view of fundamental theory as well as from a phenomenological perspective.

The simplest model of dynamical dark energy is provided by a canonical single scalar field, minimally coupled to gravity, or {\em quintessence} \citep{FerreiraJoyce1997, WangSteinhardt1998, ZlatevWangSteinhardt1999}. In general, however, Lagrangians with non-canonical kinetic terms lead as well to viable models \citep{ArmendarizPiconMukhanovSteinhardt2000}. In the absence of a compelling theory, \citet{CreminelliEtal2009} recently developed an effective theory for the perturbations in generic quintessence models\footnote{Usually this name is reserved for theories of canonical scalar fields with a potential, but we will call quintessence a generic scalar field minimally coupled to gravity.}. On cosmological scales, the relevant parameter space is two-dimensional, consisting of the equation of state parameter $w = p/\rho$ and of the speed of sound of perturbations, $c_s^2 = \del p/\del \rho$ where $p$ and $\rho$ are, respectively, the pressure and the density in the velocity orthogonal gauge. Canonical scalar fields have perturbations with $c_s = 1$, so they will not cluster appreciably on sub-horizon scales and their only observationally relevant effect is the modification of the expansion history and of the linear growth factor of dark matter perturbations.

On the other hand, the opposite case of a very small, \ie negligible, speed of sound is also very interesting. From a theoretical point of view, it is ``technically natural'', since it can be thought of as a deformation of the Ghost Condensate theory \citep{ArkaniHamedEtal2004A}, in which limit one recovers an exact shift symmetry. Moreover, if the background has $w < -1$, the only way of having a stable theory, free of ghosts and gradient instabilities, is to require a tiny speed of sound $|c_s| \lesssim 10^{-15}$ \citep{CreminelliEtal2009}. In this work we will focus on models characterized by a vanishing speed of sound: from an observational point of view, models of clustering quintessence have an important effect on structure formation, since perturbations now actively participate in the gravitational collapse and virialization of structures at low redshift, along with dark matter \citep{CreminelliEtal2010,LimSawickiVikman2010}.

Most of recent investigations on the observational effects of a clustering quintessence focused so far on the linear regime \citep{DeDeoCaldwellSteinhardt2003, WellerLewis2003, HuScranton2004, BeanDore2004, Hannestad2005, CorasanitiGiannantonioMelchiorri2005, Takada2006, Torres-RodriguezCress2007, SaponeKunz2009, SaponeKunzAmendola2010}. At the linear level, however, it is difficult to distinguish peculiar effects due to dark energy perturbations from the effects due to the background energy density and expansion history. In fact, it is reasonable to expect that the assumption of an {\em inhomogeneous} dark energy component might have specific consequences on the nonlinear evolution of structures, both at small scales, in terms of the gravitational collapse of virialized objects, as well as at larger scales, in terms of corrections to the linear solutions of large-scale structure correlation functions in the mildly nonlinear regime.

Analytical investigations on the effects of dark energy on the nonlinear evolution of structures are, at this stage, particularly relevant. In the first place, the implementation of numerical simulations for some quintessence models is particularly challenging and necessarily involves approximations. The comparison with theoretical predictions can therefore provide an essential test. In addition, for some specific models, as for instance the one considered here, numerical results are not yet available. In the second place, future large-scale structure observations will provide large data-sets with significant information in the mildly nonlinear regime: restricting our analysis to ``linear scales'' will severely limit their scientific exploitation, particularly in the case of weak lensing observations. 

For a clustering quintessence, a first step in the prediction of nonlinear effects has been taken in \citet{CreminelliEtal2010}, focusing on the extension of spherical collapse model. In the limit of zero speed of sound, pressure gradients are negligible and dark energy perturbations are comoving with the dark matter, so that the spherical halo behaves like a closed FRW universe, allowing for an exact solution. They found small modifications to the critical threshold of collapse and to the {\em dark matter} mass function. However, larger effects are expected on the {\em total} mass function, that is including the quintessence perturbation component to the virialized halos. Indeed, the relative quintessence contribution to the total halo mass is proportional to the ratio between quintessence and dark matter energy densities, {\em i.e.}~$\sim (1 + w)\,\Omega_Q/ \Omega_m$ and it is therefore particularly relevant at low redshift.

Following instead a perturbative approach to the study of density correlators in the mildly nonlinear regime, \citet{SefusattiVernizzi2011} recognized that the relevant quantity for gravitational evolution is the \emph{total} density contrast, defined as a weighted sum of dark matter and dark energy overdensities. Indeed, any measurable cosmological quantity is related gravitationally to the total density fluctuations with no possibility to distinguish in large-scale structure observations between the two components. This work shows that it is possible to write down a closed set of fluid equations for the total density and for the common velocity which take the form of the standard Eulerian equations for the dark matter, with a simple correction to the linear term in the velocity divergence of the continuity equation given by the time-dependent factor $C = 1 + (1+w) \Omega_Q/\Omega_m$. The equations of motions are solved at second order in standard Eulerian Perturbation Theory \citep[EPT, see][for a review]{BernardeauEtal2002}. The second order solution allows to derive the tree-level expression for the density bispectrum, but third order solutions are required to compute the next-to-leading order (one-loop) correction to the power spectrum. 

Other, more accurate, methods, along with standard EPT, have been proposed in recent years to predict the nonlinear matter power spectrum as well as higher-order correlation functions. As a theoretical motivation, on small scales the EPT perturbative expansion for density correlators is not well defined, because it presents large cancellations between contributions of the same order. However, it has been shown that classes of higher-order corrections can be resummed, leading to a well-established perturbative scheme known, in its first formulation, as Renormalized Perturbation Theory \citep[RPT, ][]{CrocceScoccimarro2006B, CrocceScoccimarro2006A, CrocceScoccimarro2008, BernardeauCrocceScoccimarro2008, BernardeauCrocceSefusatti2010}. Complementary methods are given by the Renormalization Group approach \citep[RG, ][]{MatarresePietroni2007, MatarresePietroni2008}, the path-integral formalism of \citet{Valageas2007}, closure theory \citep{TaruyaHiramatsu2008}, Lagrangian resummation \citep{Matsubara2008A, Matsubara2008B, Matsubara2008Berr, Matsubara2011}, and the Time-RG approach \citep[Time-RG, ][]{Pietroni2008, LesgouguesEtal2009, BartoloEtal2010}. 

In this paper, we will apply the Time-RG method to the equations of motion for the total density perturbations derived in \citet{SefusattiVernizzi2011}. Since this approach is based on differential equations {\em in time} for the matter power spectrum and bispectrum directly derived from the equations of motion of the matter and velocity field, it is particularly straightforward to adapt it to the evolution of the total density fluctuations in a model where the quintessence field has a vanishing speed of sound. In this case, in fact, the only modification to the equations of motion amounts to an extra time-dependent factor in the continuity equation. 

We will compare the nonlinear corrections to the total density power spectrum for clustering quintessence with the matter density power spectrum for $\Lambda$CDM and smooth quintessence models. We will focus in particular on the relative nonlinear corrections, that is the ratio between nonlinear and linear power spectrum, showing that the effect of a vanishing speed of sound is relatively small. However, and precisely for this reason, since at low redshift the effect on the {\em linear growth factor} of quintessence perturbations is instead particularly relevant, we will point out that the typical relation between the amplitude of linear fluctuation and the corresponding nonlinear correction, exploited by fitting formulas like \texttt{halofit} \citep{SmithEtal2003}, is not respected in the clustering quintessence model. For instance, for models with $w>-1$, the nonlinear corrections at redshift zero will be significantly smaller than those na\" ively expected from the corresponding linear power spectrum. This fact can have interesting consequences for detectability of quintessence perturbations, particularly in weak lensing observations, that will be explored in future works. 

The paper is organized as follows. In Section 2 we write down the equations of motion and discuss their linear solution. In Section 3 we describe the time-RG approach and the non-linear evolution of the power spectrum. Our results are presented in Section 4, while Section 5 is devoted to our conclusions and future prospects. In an appendix we briefly discuss the accuracy and validity of the Time-RG method.


\section{Equations of motion and linear solutions}
\label{sec:eom}

We will consider the case of a two-fluid system coupled gravitationally, following the description given by \citet{SefusattiVernizzi2011} and adopting their notation. The quintessence component is described by a constant equation of state, $w=\bar{p}_Q/\bar{\rho}_Q$ and vanishing speed of sound, $c_{s}=0$. Under these assumptions, the matter and quintessence perturbations are comoving and therefore share the same velocity field. The relevant degrees of freedom are the matter and quintessence density contrasts, respectively $\d_m$ and $\d_Q$, and their common velocity $\vv$. The equations of motion are given by two continuity equations corresponding to the two components,
\begin{align}
&\frac{\partial \d_m}{\partial \tau}  + \vec \nabla \cdot \big[ (1+\d_m) \vv \big] = 0 \label{continuity_m}\,,\\
&\frac{\partial \delta_Q}{\partial \tau}  -3 w \HH \delta_Q+ \vec \nabla \cdot \big[ (1+w+\delta_Q) \vv\big] = 0 \label{continuity_Q}\,,
\end{align}
and one Euler equation, 
\be
\frac{\partial {\vec{ v}}}{\partial \tau}  + \HH \vec v+ (\vec v \cdot \vec \nabla) \vec v =-\vec \nabla \Phi\;, \label{euler_common}
\ee
where $\tau$ is the conformal time and $\HH=d\ln a/d \tau$. In addition, we consider the Poisson equation relating the energy density to the gravitational potential $\Phi$,
\be
\nabla^2 \Phi = \frac{3}{2}\, \HH^2\, \Omega_m \left( \delta_m +  \delta_Q \frac{\Omega_Q}{\Omega_m} \right)\,,
\ee
where the relative mean densities of both matter and quintessence are given by $\Omega_\alpha=\Omega_\alpha(\tau)\equiv\bar{\rho}_\alpha/(\bar{\rho}_m+\bar{\rho}_Q)$ with $\alpha=m$, $Q$.
As customary, we define the velocity divergence $\theta\equiv\vec \nabla \cdot \vec v$, fully describing the peculiar velocity $\vec v$ up to shell crossing \citep{BernardeauEtal2002}. In Fourier space, assuming the notations $q_i \equiv | \vec q_i|$ and $\vec q_{ij}\equiv\vec q_i+\vec q_j$, the continuity equations become 
\begin{align}
&\frac{\partial \delta_{m,\kv}}{\partial \tau}  + \theta_{\kv} = -\!\!\int\!\! d^3 q_1 d^3 q_2\, \delta_D (\vec k - \vec q_{12})\,\alpha(\qv_1, \qv_2)\, \theta_{\vec q_1} \delta_{m,\vec q_2} \label{continuity_m_FS}\,,\\
&\frac{\partial \delta_{Q,\vec k}}{\partial \tau}  -3 w \HH \delta_{Q,\vec k}+ (1+w) \theta_{\vec k} = -\!\!\int\!\! d^3 q_1 d^3 q_2\, \delta_D (\vec k - \vec q_{12})\,\alpha(\vec q_1, \vec q_2)\, \theta_{\vec q_1} \delta_{Q,\vec q_2} \label{continuity_Q_FS}\,,
\end{align}
with
\be
\alpha(\vec q_1, \vec q_2)  \equiv  1 + \frac{\vec q_1 \cdot \vec q_2}{q_1^2}  \;.  \label{alpha_def}
\ee
The Poisson equation can be used to replace the gravitational potential in the Euler equation to obtain for the latter, in Fourier space
\be
\frac{\partial \theta_{\kv}}{\partial \tau} + \HH\, \theta_{\kv} + \frac32\, \Omega_m \HH^2 \left( \delta_{m,\kv} + \delta_{Q,\kv}\, \frac{\Omega_Q}{\Omega_m} \right) = -\!\!\int\!\! d^3 q_1 d^3 q_2\, \delta_D (\vec k - \vec q_{12})\,\beta(\vec q_1, \vec q_2)\, \theta_{\vec q_1} \,\theta_{\vec q_2}\;, \label{euler_sum}
\ee
with
\be
\beta(\vec q_1, \vec q_2)\equiv \frac{q_{12}^2 \,\vec q_1 \cdot \vec q_2 }{2\, q_1^2\, q_2^2}\;.\label{beta_def}
\ee
As pointed out by \citet{SefusattiVernizzi2011}, it is convenient to combine matter and quintessence perturbations into a {\em total density contrast} defined as
\be \label{tdc}
\delta \equiv \frac{\delta \rho}{\bar \rho_m} = \delta_m +  \delta_Q \frac{\Omega_Q}{\Omega_m}\;,
\ee
which is bounded from below as $\d>-1-\Omega_Q/\Omega_m$, instead of the usual $\d>-1$.
In fact, any clustering observable is affected only by the gravitational effects of the total energy density, while the individual matter and quintessence components are {\em not}---even indirectly---accessible by observations. The gravitational potential, responsible for both weak lensing effects and galaxy clustering is given, in terms of the total density perturbations $\d$, by
\be 
\nabla^2 \Phi = 4 \pi\, G\, a^2 \bar \rho_m\, \delta \;. \label{poisson_2}
\ee
For this reason, we will focus in this work on the nonlinear evolution of the total density perturbations defined by eq.~(\ref{tdc}). Clearly, other definitions of such quantity, differing by an overall time-dependent factor, are possible. This definition has the advantage of recovering standard expressions when quintessence perturbations are not present. Introducing the quantity
\be\label{eq:C}
C(\tau) \equiv  1+ (1+w)\frac{\Omega_Q(\tau)}{\Omega_m(\tau)}\;,
\ee
we can rewrite the equations of motion for the total density contrast $\d$ and the velocity divergence $\theta$ as  
\begin{align}
&\frac{\partial \delta_{\vec k}}{\partial \tau}  + C  \theta_{\vec k} = -\!\!\int\!\! d^3 q_1 d^3 q_2\, \delta_D (\vec k - \vec q_{12})\,\alpha(\vec q_1, \vec q_2) \theta_{\vec q_1} \delta_{\vec q_2} \label{continuity_tot}\,,\\
&\frac{\partial \theta_{\vec k}}{\partial \tau} + \HH \theta_{\vec k} + \frac{3}{2} \Omega_m \HH^2 \delta_{\vec k} = - \!\!\int\!\! d^3 q_1 d^3 q_2\, \delta_D (\vec k - \vec q_{12})\,\beta(\vec q_1, \vec q_2) \theta_{\vec q_1} \theta_{\vec q_2}\;. \label{euler_tot}
\end{align}
Notice that setting $C=1$ reduces the system to the usual smooth quintessence scenario with $\d=\d_m$. On the other hand, the effect of quintessence perturbations is expected to be relevant at late times and for significant departures from $w=-1$, with a sign depending on the sign of $1+w$ \citep{WellerLewis2003, CreminelliEtal2009}. 

\begin{figure}[t]
\begin{center}
{\includegraphics[width=0.49\textwidth]{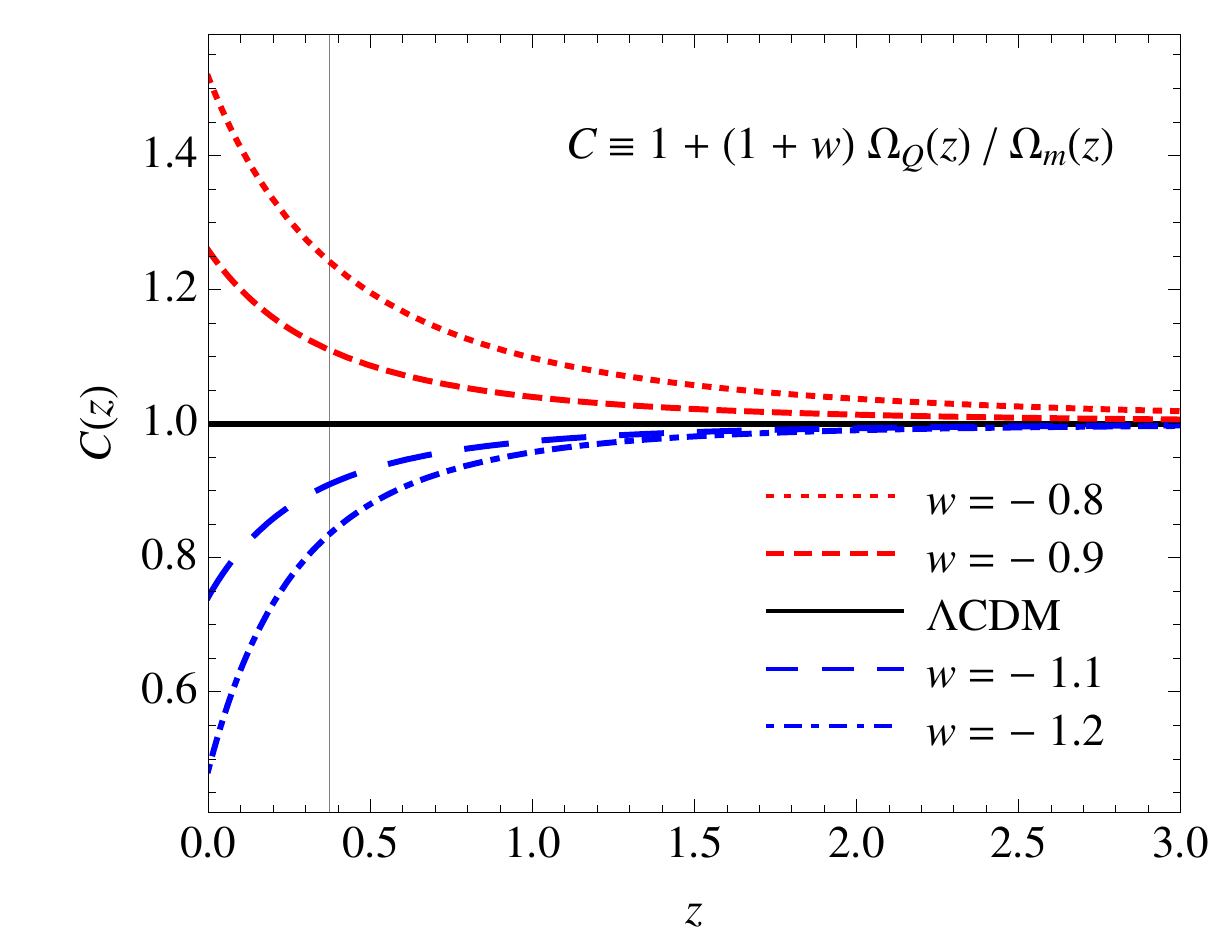}}
{\includegraphics[width=0.49\textwidth]{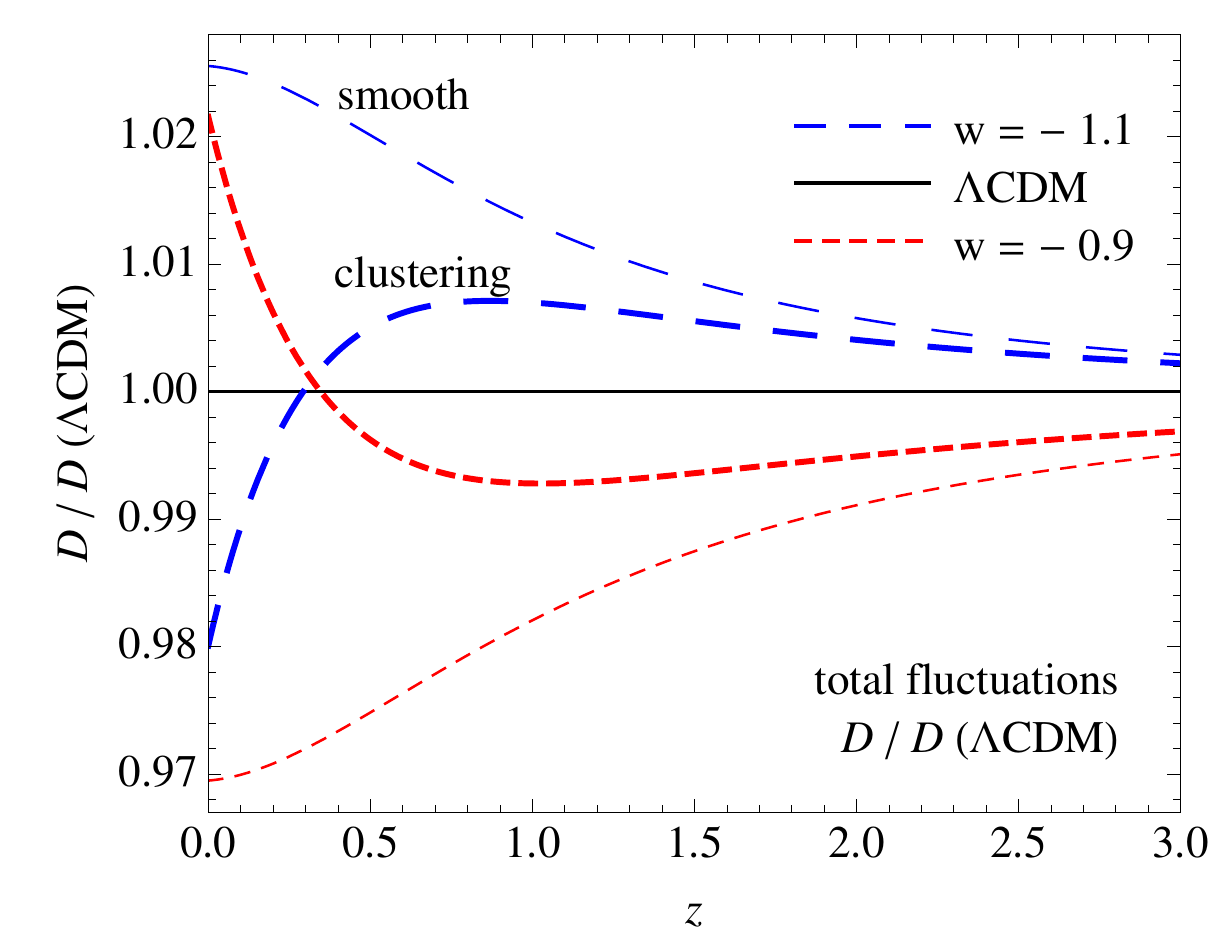}}
\caption{{\em Left panel}: The function $C(z)$, eq.~(\ref{eq:C}), as a function of redshift for $w=-1.2$ ({\em blue, dot-dashed curves}), $w=-1.1$ ({\em blue, long-dashed curves}), $w=-1$ ($\Lambda$CDM, {\em black, continuous curves}), $w=-0.9$ ({\em red, short-dashed curves}) and $w=-0.8$ ({\em red, dotted curves}). The vertical line indicates the redshift of equality between matter and quintessence for a $\Lambda$CDM cosmology, \ie~$z_{eq}=0.37$ for the assumed $\Omega_{m,0}=0.279$. {\em Right panel}: ratio of linear growth factor $D$ to the same quantity in the $\Lambda$CDM model as a function of redshift, assuming $w=-1.1$ ({\em blue, long-dashed curves}) and $w=-0.9$ ({\em red, short-dashed curves}), with thick and thin curves corresponding respectively to the clustering and smooth cases. The two panels reproduce Fig.~1 and the bottom left panel of Fig.~2 of \citet{SefusattiVernizzi2011}.}
\label{fig:CD}
\end{center}
\end{figure}
Since the function $C(z)$ is uniquely responsible for the correction to the equations of motion induced by the clustering of quintessence, and therefore particularly relevant to our problem, we plot it as a function of redshift for a few values of the equation of state $w$ in the left panel of Fig.~\ref{fig:CD}, reproducing Fig.~1 in \citet{SefusattiVernizzi2011}. In particular we consider the following values and corresponding graphic notations, assumed throughout the paper, for the quintessence equation of state: $w=-1.2$ ({\em blue, dot-dashed curves}), $w=-1.1$ ({\em blue, long-dashed curves}), $w=-1$ ($\Lambda$CDM, {\em black, continuous curves}), $w=-0.9$ ({\em red, short-dashed curves}) and $w=-0.8$ ({\em red, dotted curves}). In addition, when needed, thin lines indicate smooth quintessence results as opposed to thick lines corresponding to the clustering case. Here and in the rest of the paper we assume a flat Universe with $\Omega_{m,0}=0.279$, ``0'' denoting quantities at $z=0$. In the $\Lambda$CDM scenario, this implies a redshift of equality between matter and cosmological constant components equal to $z_{eq}\simeq 0.37$, shown by the vertical line in Fig.~\ref{fig:CD}. For all results involving the initial power spectrum in the next sections we further assume $\Omega_b=0.046$, $h=0.7$, $n_s=0.96$ and $\sigma_8=0.8$ for the $\Lambda$CDM case. We fixed the normalization of the primordial power spectrum to be the same for all models, so that the actual value of $\sigma_8$ for models different from $\Lambda$CDM in fact depends on the value of the linear growth factor at redshift zero.

The linear solutions to the equations of motion of matter and quintessence perturbations and their observational consequences have been studied in several works \citep{DeDeoCaldwellSteinhardt2003, WellerLewis2003, BeanDore2004, HuScranton2004, CorasanitiGiannantonioMelchiorri2005, Hannestad2005, Takada2006, Torres-RodriguezCress2007, SaponeKunz2009, SaponeKunzAmendola2010}. In particular, \citet{SefusattiVernizzi2011} find, for the total overdensity $\d$, an integral solution to the linear growth factor $D(z)$ completely analogous to the one obtained in $\Lambda$CDM cosmologies (but absent in smooth quintessence scenarios), given by
\be
D(a) = \frac52\, H_0^2\, \Omega_{m,0}\, H(a)\! \int_0^a\!\! \frac{C(\tilde a)}{[\tilde a\, H(\tilde a)]^3} d \,\tilde a\;. \label{int_sol}
\ee
The corresponding linear growth factor for matter perturbations alone is given by
\be
D_m(a)=\int_0^a\left[\frac52 -\frac32\frac{D(\tilde{a})}{\tilde{a}}\right]\Omega_m(\tilde{a})\,d\, \tilde{a}\;,
\ee
but we will not consider the evolution of matter perturbations alone when quintessence perturbations are present, focusing instead on the total density fluctuations throughout the paper.  

In order to make this work as much self-contained as possible and to help in the interpretations of our results in section~\ref{sec:results}, we show, in the right panel of Fig.~\ref{fig:CD}, the ratio of linear growth factor $D$ to the same quantity in the $\Lambda$CDM model as a function of redshift, assuming $w=-1.1$ ({\em blue, long-dashed curves}) and $w=-0.9$ ({\em red, short-dashed curves}), with thick and thin curves corresponding respectively to the clustering and smooth cases. It is crucial to remark that while the high-redshift behavior is similar and asymptotically equal in the $c_s=0$ as in the $c_s=1$ scenarios, the growth of quintessence perturbations strongly affects the low-$z$ behavior leading to a turn-around and larger or smaller values for the growth factor w.r.t. $\Lambda$CDM, corresponding, respectively to $w=-0.9$ or $w=-1.1$. For a detailed discussion of the phenomenology of the linear growth of total density fluctuations see \citet{SefusattiVernizzi2011}.


\section{Non-linear evolution and the Time-RG approach}
\label{sec:nonlinear}

In this section we study the nonlinear evolution of the total density perturbations in terms of the nonlinear corrections to the total density power spectrum. In this respect, as we will see, the Time-Renormalization Group approach of \citet{Pietroni2008} is particularly apt at extending the linear theory predictions, due to the specific form of the correction induced by clustering quintessence on the equations of motion. 

We use, as time variable, $\eta\equiv \ln D$ with $D$ corresponding to the growing mode, and rescale the velocity divergence $\theta$ in the following way:
\be
\tt \equiv  - \frac{C }{\HH f} \theta \;, \label{Theta_def}
\ee
where $f\equiv d\ln D/ d\ln a$ is the linear growth rate. The linear solutions for the density and velocity are then given by
\be
\tt^{\rm lin}_{\vec k} (\eta) =  \d_{\vec k}^{\rm lin} (\eta)= D (\eta) \delta^{\rm in}_{\vec k} \;,  \label{theta_lin}
\ee
$\delta^{\rm in}_{\vec k}$ corresponding to the initial conditions.
In the new variables, the equations of motion become
\begin{align}
&\frac{\partial \delta_{\vec k}}{\partial \eta} -   \tt_{\vec k} = \!\!\int\!\! d^3 q_1 d^3 q_2\, \delta_D (\vec k - \vec q_{12})\,\frac{\alpha(\vec q_1, \vec q_2)}C \tt_{\vec q_1} \delta_{\vec q_2} \label{continuity_tot_eta}\,,\\
&\frac{\partial \tt_{\vec k}}{\partial \eta} - \tt_{\vec k} +\frac32 \frac{\Omega_m\,C}{f^2} (\tt_{\vec k} - \delta_{\vec k}) = \!\!\int\!\! d^3 q_1 d^3 q_2\, \delta_D (\vec k - \vec q_{12})\, \frac{\beta(\vec q_1, \vec q_2)}C \tt_{\vec q_1} \tt_{\vec q_2}\;. \label{euler_tot_eta}
\end{align}
Again, the only difference with the standard equations of motions for matter perturbations in the $\Lambda$CDM or smooth quintessence scenarios is given by the presence of the function $C(z)$, different from unity at small redshift. 

\citet{SefusattiVernizzi2011} studied the nonlinear solutions to these equations of motion in the standard Eulerian Perturbation Theory (EPT) framework, finding the second-order solution for the density and velocity fields.
Such corrections are sufficient to compute the tree-level expression for the total density bispectrum, but the first nonlinear (one-loop) correction to the power spectrum requires the knowledge of the third order solution for $\d_{\vec k}$ and $\tt_{\vec k}$.

In this work we will follow, however, a different approach. The idea at the basis of the Time-RG method is to take ensemble averages of the continuity and Euler equations, obtaining a hierarchy of differential equations in time for the density and velocity correlation functions \citep{Pietroni2008}. Because of nonlinearities, the equations for the $n$-th order correlation will involve the $(n+1)$-th order one, so one has to truncate the hierarchy at some order: since we are interested in the non-linear power spectrum, it has been shown that a good approximation is given by setting the connected 4-point function, or {\em trispectrum}, to zero.

It is convenient to rewrite the fluid equations in terms of the doublet \citep{Scoccimarro1998}
\be
\begin{pmatrix}
\phi_{1, \kv}(\eta) \\
\phi_{2, \kv}(\eta)
\end{pmatrix} \equiv
e^{-\eta} 
\begin{pmatrix}
\d_{\kv}(\eta) \\
\tt_{\kv}(\eta)
\end{pmatrix} \; ,
\ee
so that eqs.~\eqref{continuity_tot_eta}-\eqref{euler_tot_eta} become
\be
\label{eq:master}
\de_\eta \phi_{a, \kv}(\eta) + \Om_{ab}(\eta)\,\phi_{b, \kv}(\eta)
= \frac{e^\eta}{C(\eta)}\,\gam_{abc}(\kv; -\qv_{1},-\qv_{2})\, \phi_{b, \qv_1}(\eta) \phi_{c, \qv_2}(\eta) \; ,
\ee
where we have defined the time-dependent, inverse``propagator''
\be
\Om_{ab}(\eta) \equiv
\begin{pmatrix}
1 & -1 \\
-\frac{3}{2}\,\frac{\Om_{m}\,C}{f^2} & \frac{3}{2}\,\frac{ \Om_{m}\,C}{f^2}
\end{pmatrix}\;,
\ee
and the vertex function is given by
\begin{gather}
\gam_{121}(\kv; \qv_1, \qv_2) = \gam_{112}(\kv; \qv_2, \qv_1) = \d_D(\kv+\qv_{12})\, \alpha(\qv_1, \qv_2)/2 \\
\gam_{222}(\kv; \qv_1, \qv_2) = \d_D(\kv+\qv_{12})\, \bt(\qv_1, \qv_2) \; ,
\end{gather}
with all other components being equal to zero.

Following \citet{Pietroni2008}, we now consider derivatives of the correlation functions, and using eq.~\eqref{eq:master} we find the Time-RG hierarchy
\begin{align}
\de_\eta \avg{\phi_a \phi_b} =& -\Om_{ac} \avg{\phi_c \phi_b} - \Om_{bc} \avg{\phi_c \phi_a}
+ \frac{e^\eta}{C} \gam_{acd} \avg{\phi_c \phi_d \phi_b} + \frac{e^\eta}{C} \gam_{bcd} \avg{\phi_c \phi_d \phi_a} \\
\de_\eta \avg{\phi_a \phi_b \phi_c} =& -\Om_{ad} \avg{\phi_a \phi_ b \phi_c} - \Om_{bd} \avg{\phi_a \phi_d \phi_c}
-\Om_{cd} \avg{\phi_a \phi_b \phi_d} \nonumber \\ 
&+ \frac{e^\eta}{C} \gam_{ade} \avg{\phi_d \phi_e \phi_b \phi_c} + \frac{e^\eta}{C} \gam_{bde} \avg{\phi_a \phi_d \phi_e \phi_c}
+ \frac{e^\eta}{C} \gam_{cde} \avg{\phi_a \phi_b \phi_d \phi_e} \\
\de_\eta \avg{\phi_a \phi_b \phi_c \phi_d} =& \dots
\end{align}
We truncate the hierarchy setting $\avg{\phi_a \phi_b \phi_c \phi_d}_c = 0$ and, using the definitions of the power spectrum and bispectrum
\begin{align}
&\avg{\phi_{a, \kv_i} \phi_{b, \kv_j}} = \d_D(\kv_{ij}) P_{ab}(k_i) \; , \\
&\avg{\phi_{a, \kv_i} \phi_{b, \kv_j} \phi_{c, \kv_l}} = \d_D(\kv_{ijl}) B_{abc}(k_i,k_j,k_l) \; ,
\end{align}
our system of equations can be rewritten as
\begin{align}
&\de_\eta P_{ab}(k) = -\Om_{ac} P_{cb}(k) - \Om_{bc} P_{ac}(k)
+ \frac{e^\eta}{C} \frac{2 \pi}{k} \left[ I_{acd, bcd}(k) + I_{bcd, acd}(k) \right] \\
&\de_\eta I_{acd,bef}(k) = - \Om_{bg} I_{acd,gef}(k) - \Om_{eg} I_{acd,bgf}(k) - \Om_{fg} I_{acd,beg}(k)
+ 2 \frac{e^\eta}{C} A_{acd,bef}(k) \; ,
\end{align}
where we have introduced, for convenience, the following functions:
\begin{gather}
I_{acd,bef}(k) \equiv \int_0^{\infty} \!\!\rmd q\, q\, \int_{|q-k|}^{q+k}\!\! \rmd p\, p\, \frac{1}{2}\, \left[ \gam_{acd}(k,q,p) B_{bef}(k,q,p)
+ (q \leftrightarrow p) \right] \; , \\
A_{acd,bef}(k) \equiv \int_0^{\infty}\!\! \rmd q\, q\, \int_{|q-k|}^{q+k}\!\! \rmd p\, p\, \frac{1}{2}\, \bigg\{ \gam_{acd}(k,q,p)
\big[ \gam_{bgh}(k,q,p) P_{ge}(q) P_{hf}(p) + \gam_{egh}(q,p,k) P_{gf}(p) P_{hb}(k) \nonumber \\
+ \gam_{fgh}(p,k,q) P_{gb}(k) P_{he}(q) \big] + (q \leftrightarrow p) \bigg\} \; .
\end{gather}
The derivation is therefore the same as the one of \citet{Pietroni2008} for the standard case of matter perturbations, with the only difference given by the function $C(\eta)\ne 1$} accounting for the total density field. 

Having specified the dynamics, we need to set up the initial conditions. We start the numerical evaluations at the initial redshift $z_{in} = 50$ and we choose $P_{ab}(k,\eta_{in})$ as the linearly evolved power spectrum, while we set $B_{abc}(k,\eta_{in}) = 0$ for Gaussian initial conditions.


\section{Results}
\label{sec:results}

In this section we present our results for the nonlinear power spectrum of the total density perturbations, both in smooth and clustering quintessence cosmologies, computed in the TRG approach. 

\begin{figure}[t!]
\begin{center}
{\includegraphics[width=0.49\textwidth]{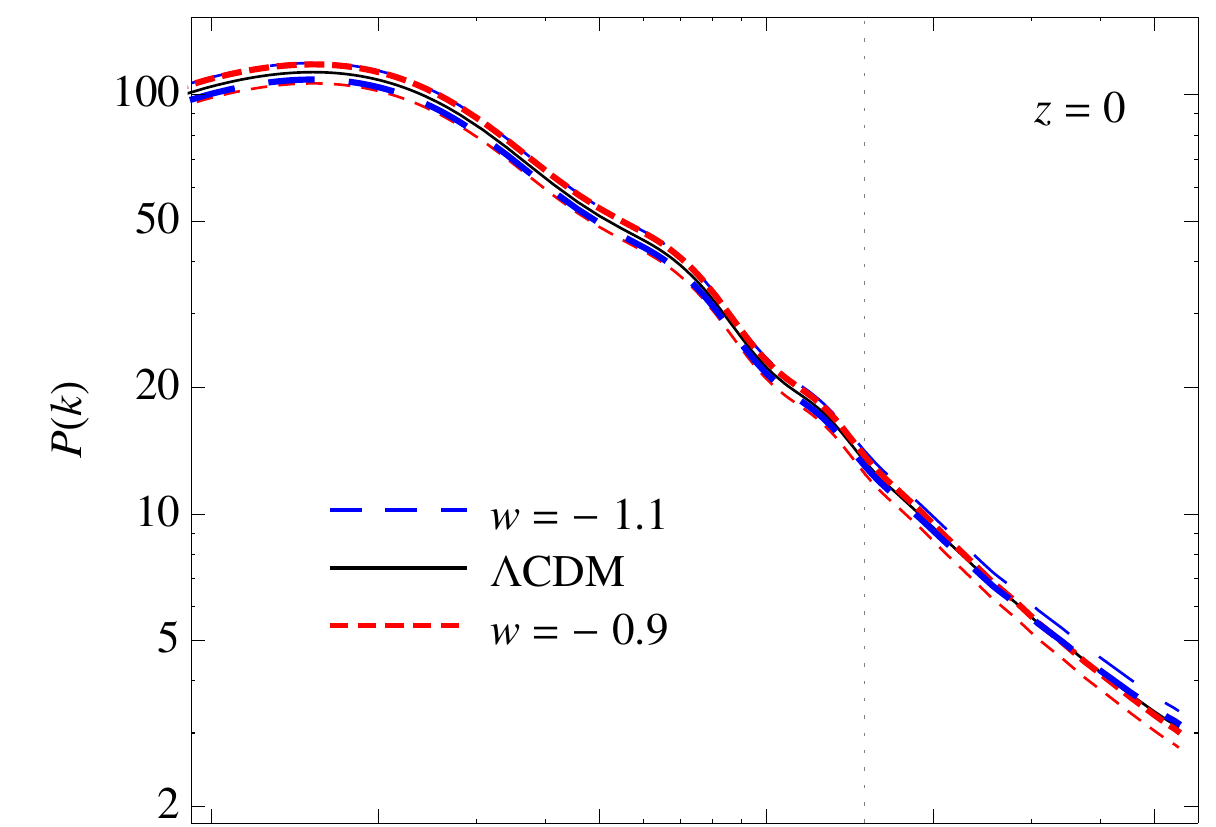}}
{\includegraphics[width=0.49\textwidth]{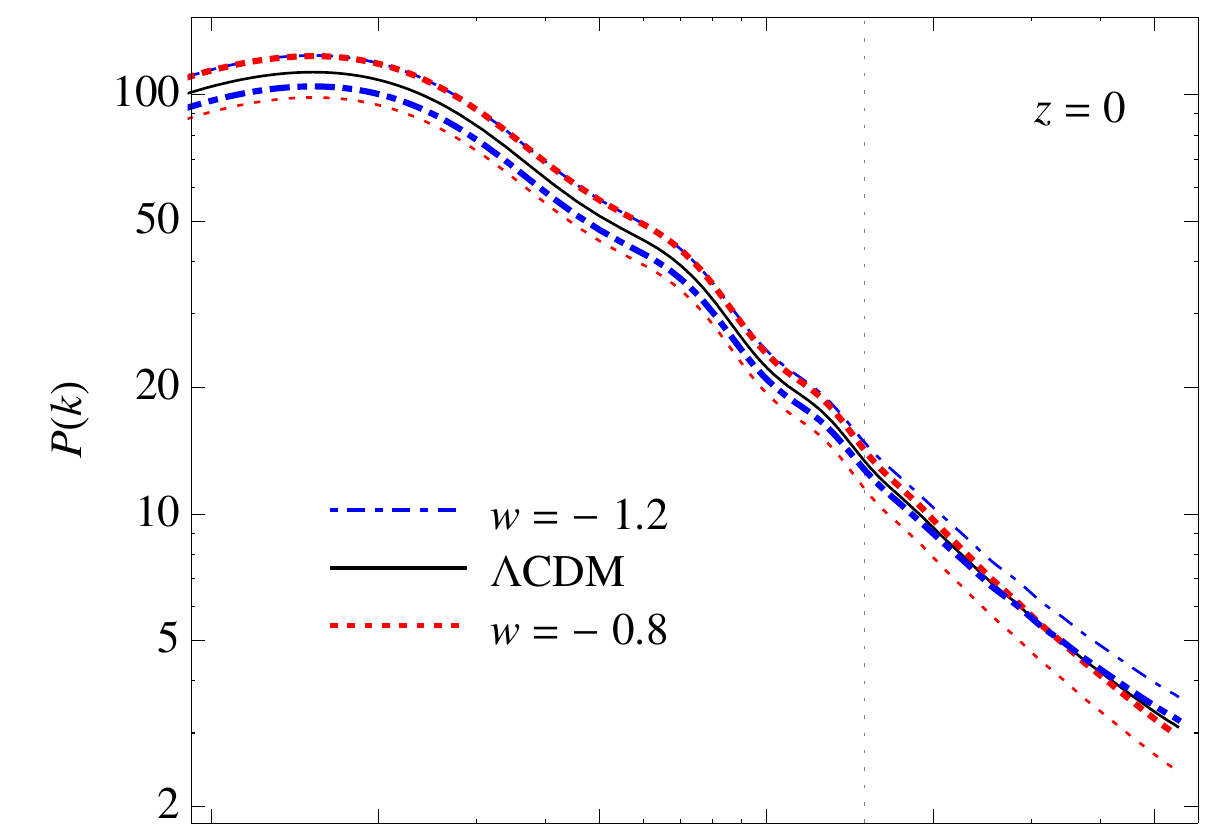}}\\
{\includegraphics[width=0.49\textwidth]{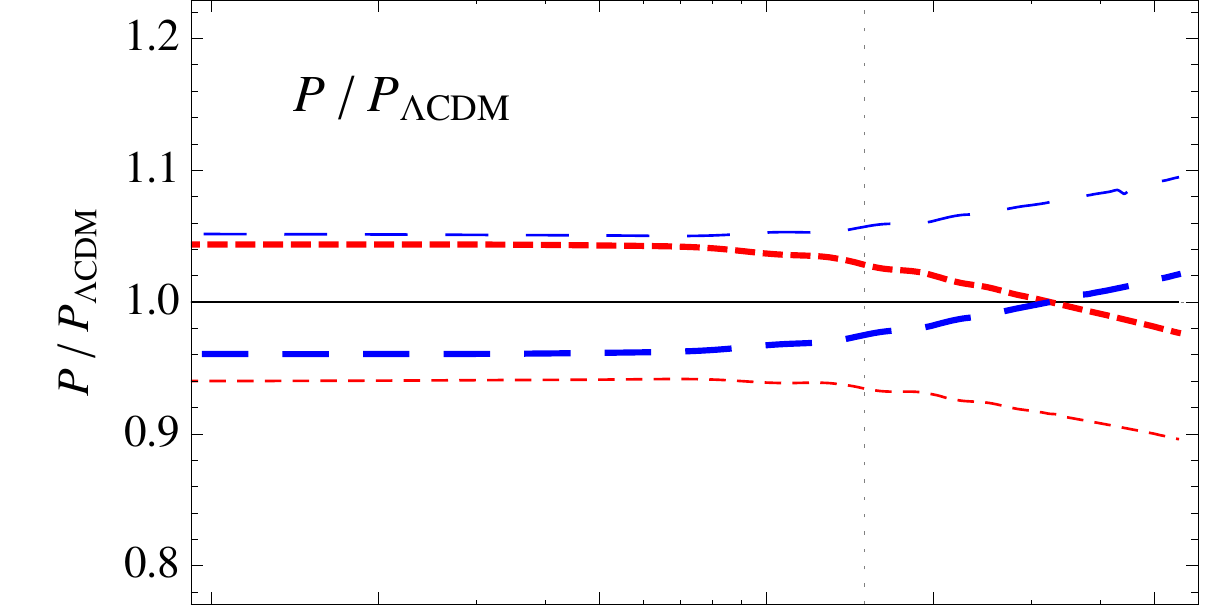}}
{\includegraphics[width=0.49\textwidth]{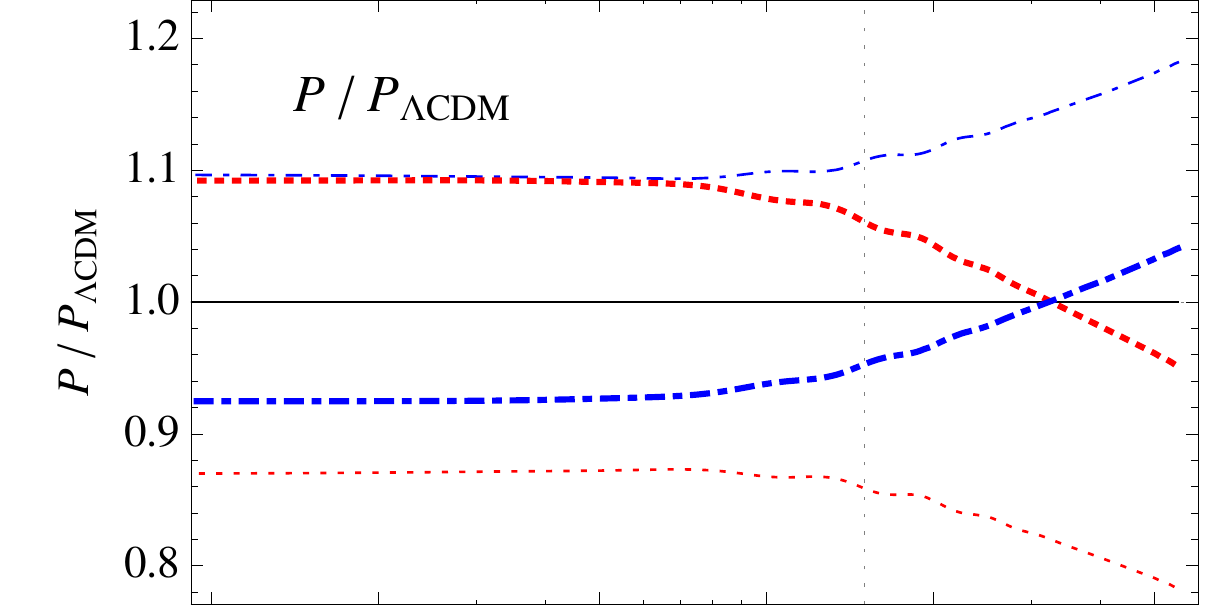}}\\
{\includegraphics[width=0.49\textwidth]{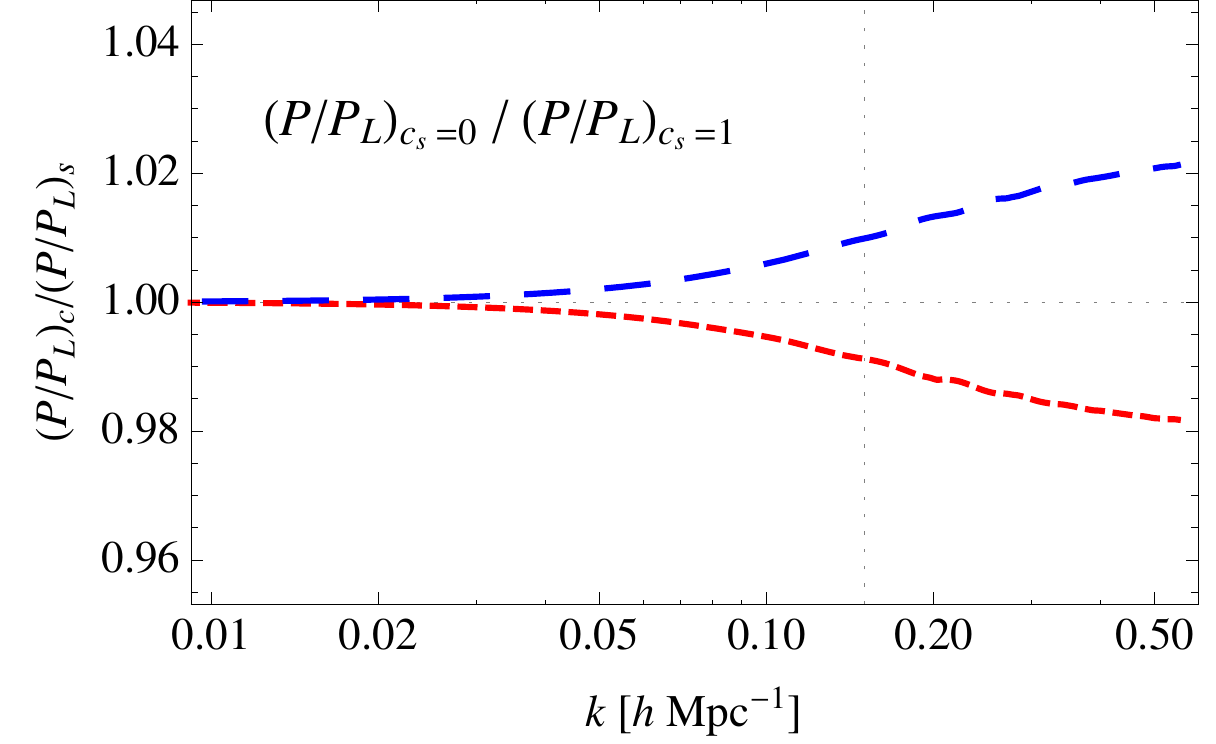}}
{\includegraphics[width=0.49\textwidth]{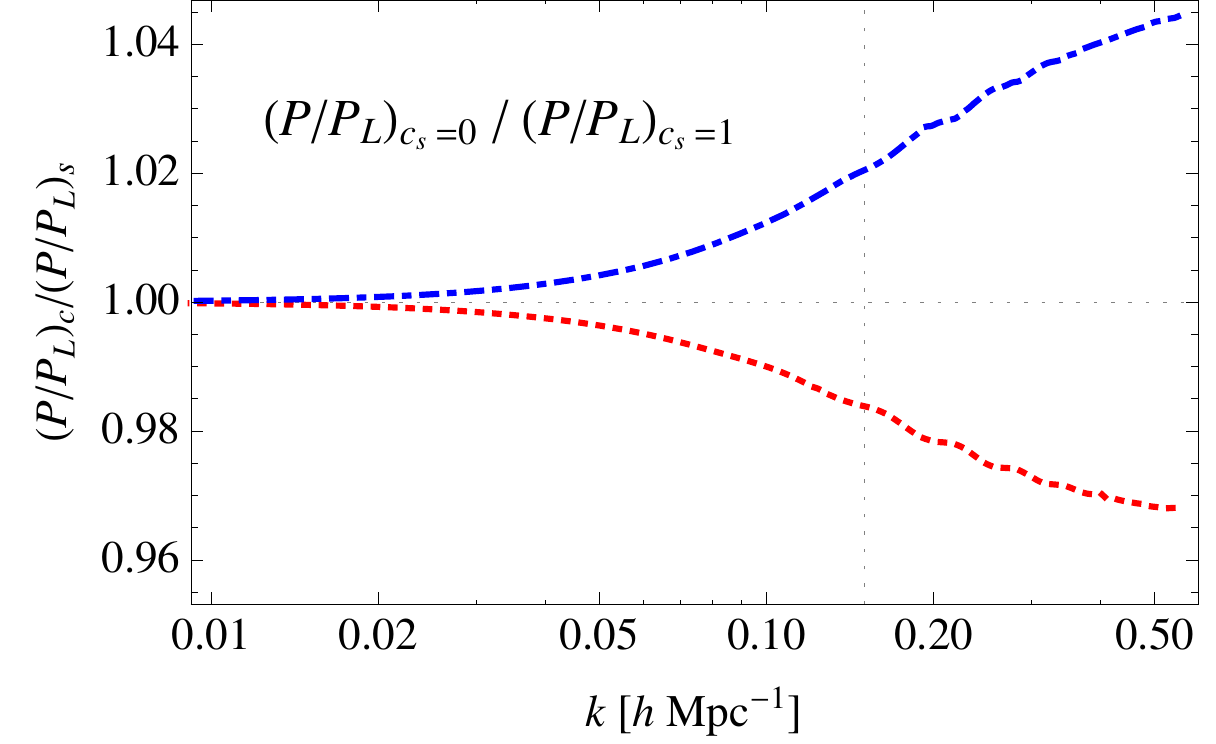}}
\caption{{\em Upper panels}: nonlinear power spectrum at $z=0$ in the $\Lambda$CDM cosmology and for $w=-0.9$ and $-1.1$ ({\em left}) and $w=-0.8$ and $-1.2$ ({\em right}). Thick lines correspond to the clustering case while thin lines correspond to the smooth case. {\em Middle panels}: ratios to the $\Lambda$CDM nonlinear power spectrum. {\em Lower panels}: ratio of the nonlinear to linear power spectrum in the clustering case divided by the same ratio for the smooth case, illustrating the effect of clustering quintessence on the nonlinear corrections. The dotted vertical line indicates the value of the wavenumber below which the TRG prediction is accurate at 1\% level in the $\Lambda$CDM case, that is $k=0.15\kMpc$.}
\label{fig:Pz0}
\end{center}
\end{figure}
\begin{figure}[t]
\begin{center}
{\includegraphics[width=0.49\textwidth]{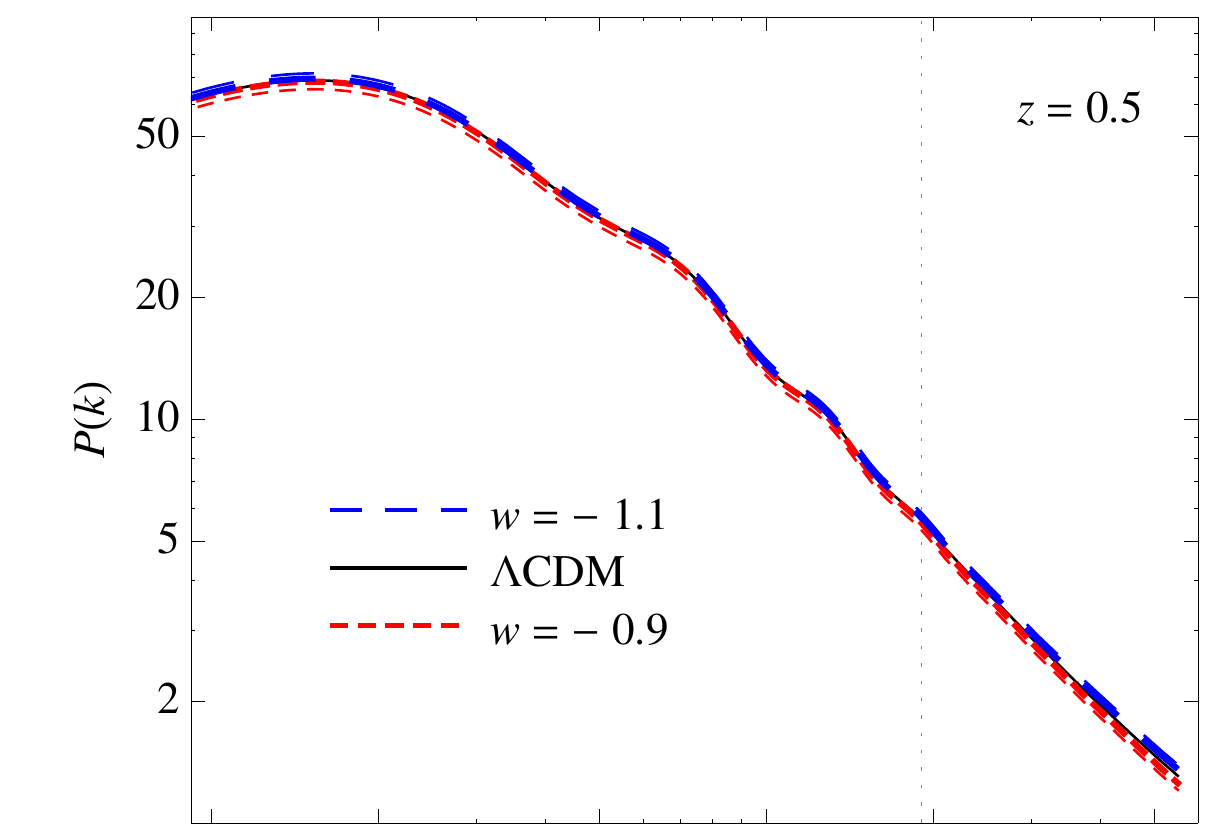}}
{\includegraphics[width=0.49\textwidth]{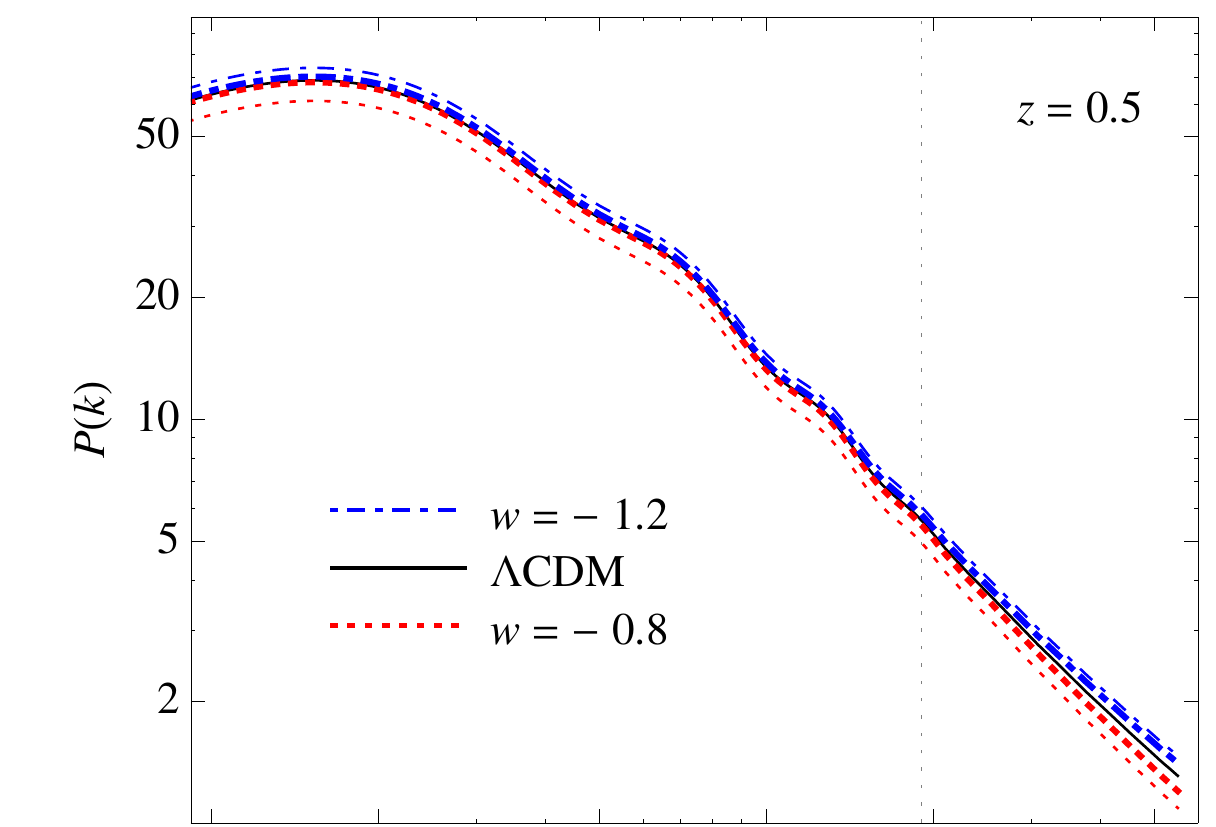}}\\
{\includegraphics[width=0.49\textwidth]{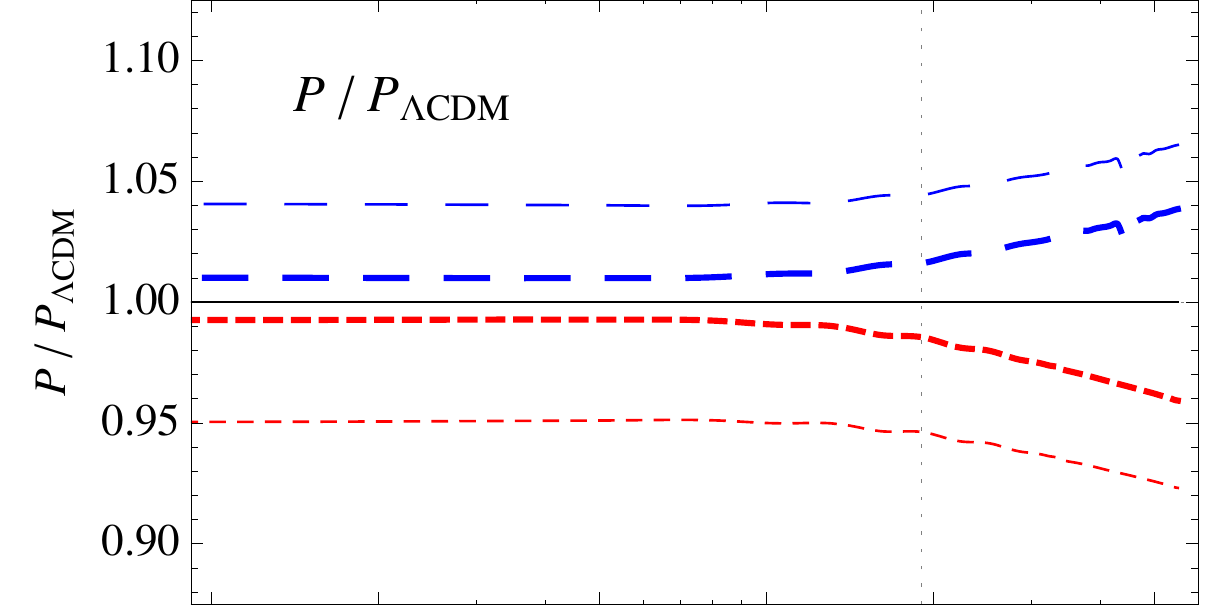}}
{\includegraphics[width=0.49\textwidth]{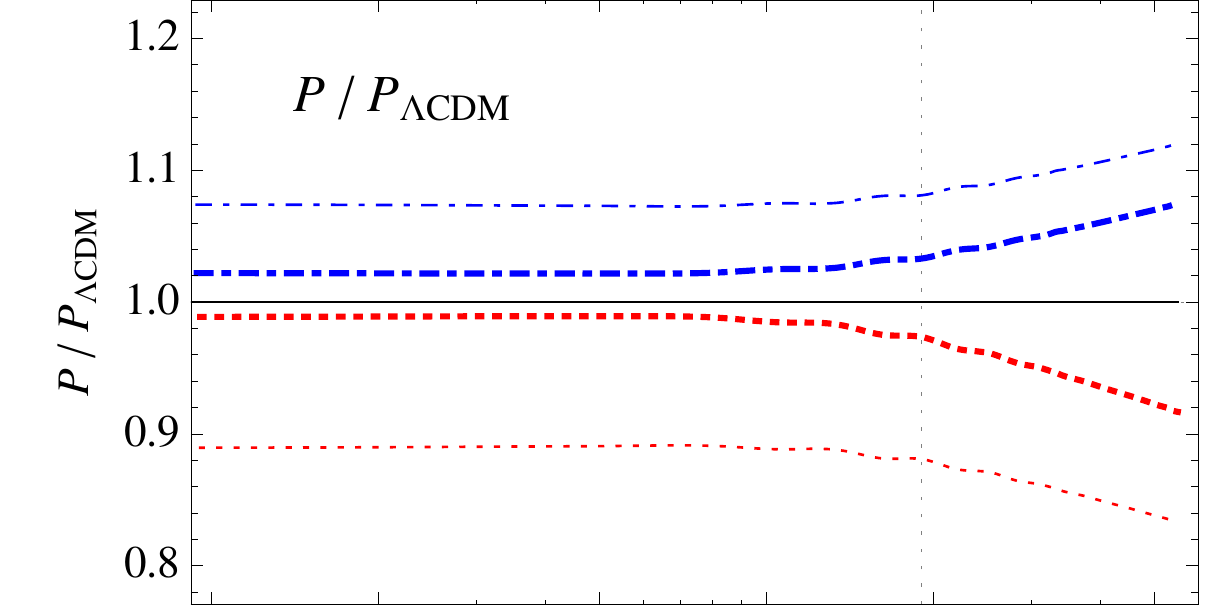}}\\
{\includegraphics[width=0.49\textwidth]{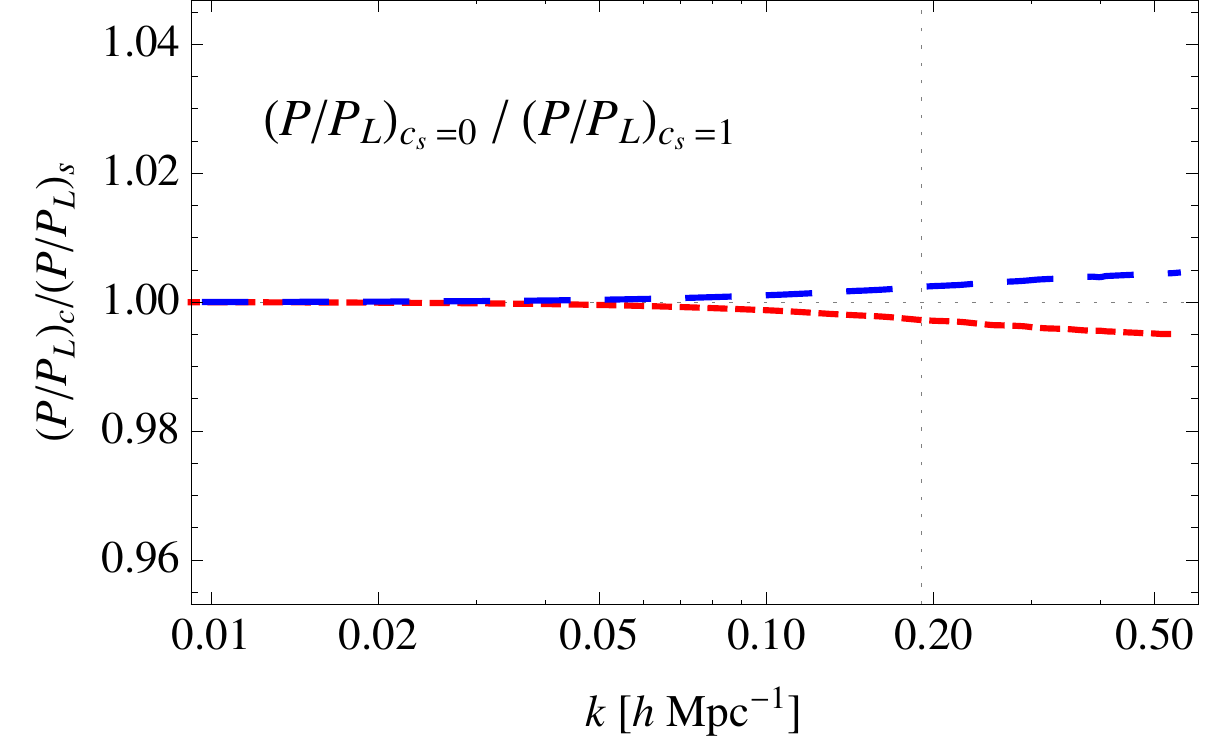}}
{\includegraphics[width=0.49\textwidth]{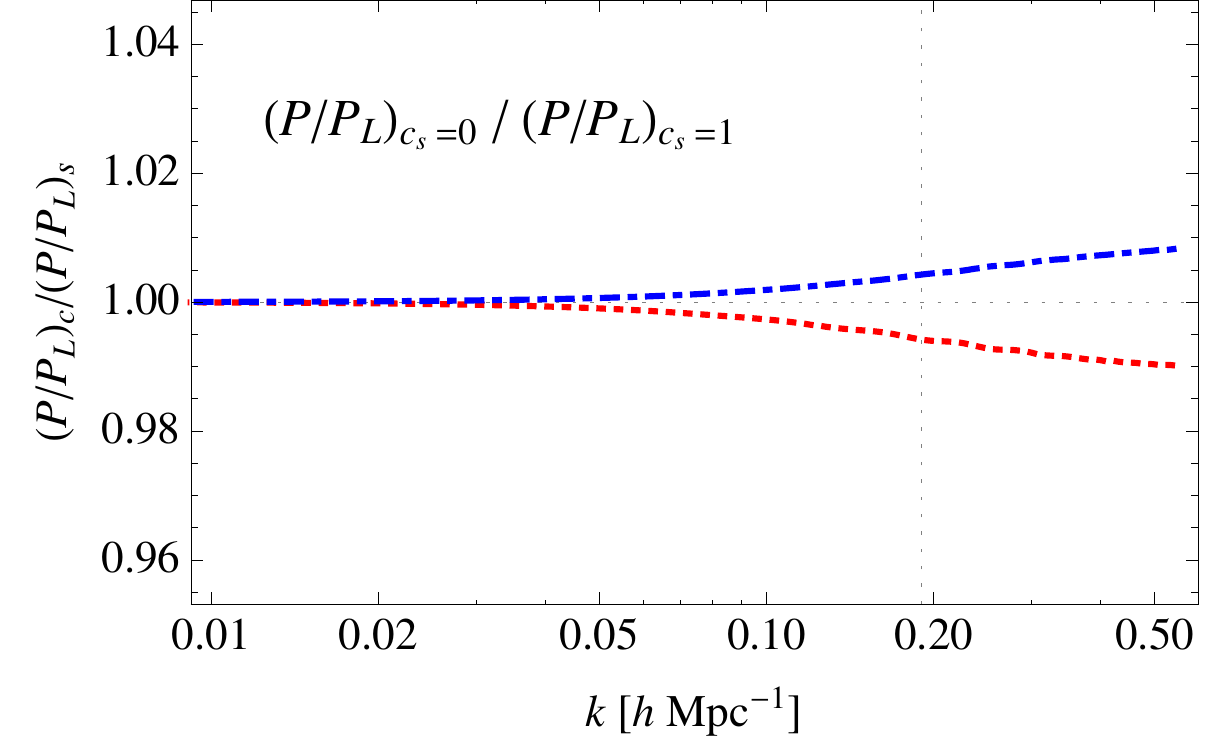}}
\caption{Same as Fig.~\ref{fig:Pz0} but at redshift $z=0.5$.}
\label{fig:Pz0p5}
\end{center}
\end{figure}
\begin{figure}[t]
\begin{center}
{\includegraphics[width=0.49\textwidth]{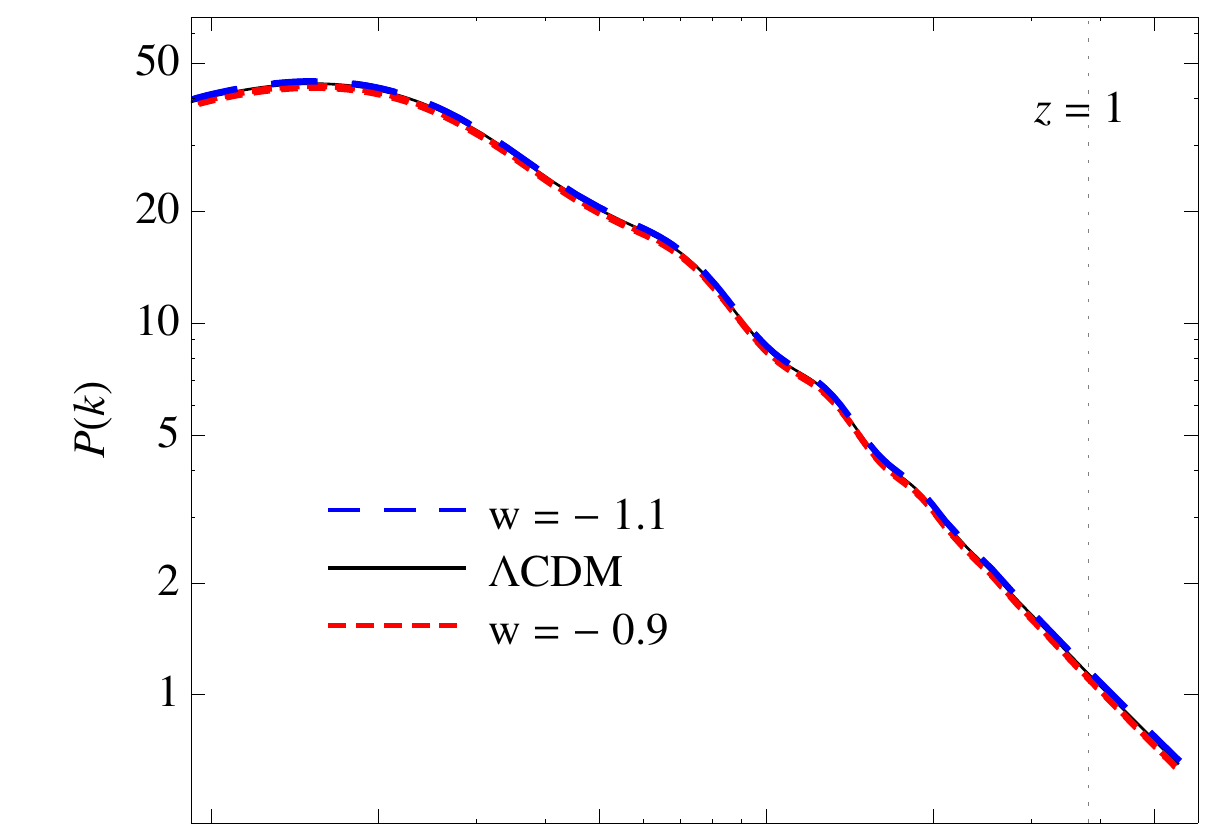}}
{\includegraphics[width=0.49\textwidth]{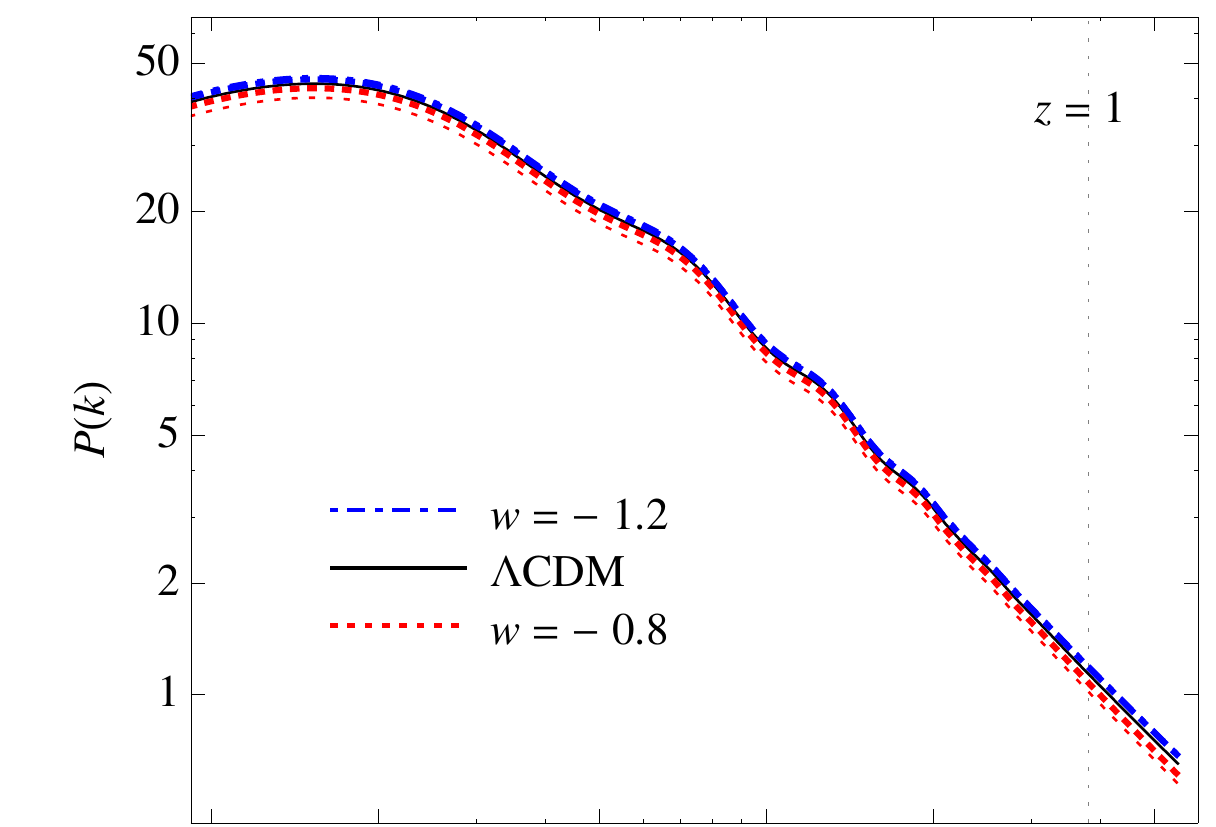}}\\
{\includegraphics[width=0.49\textwidth]{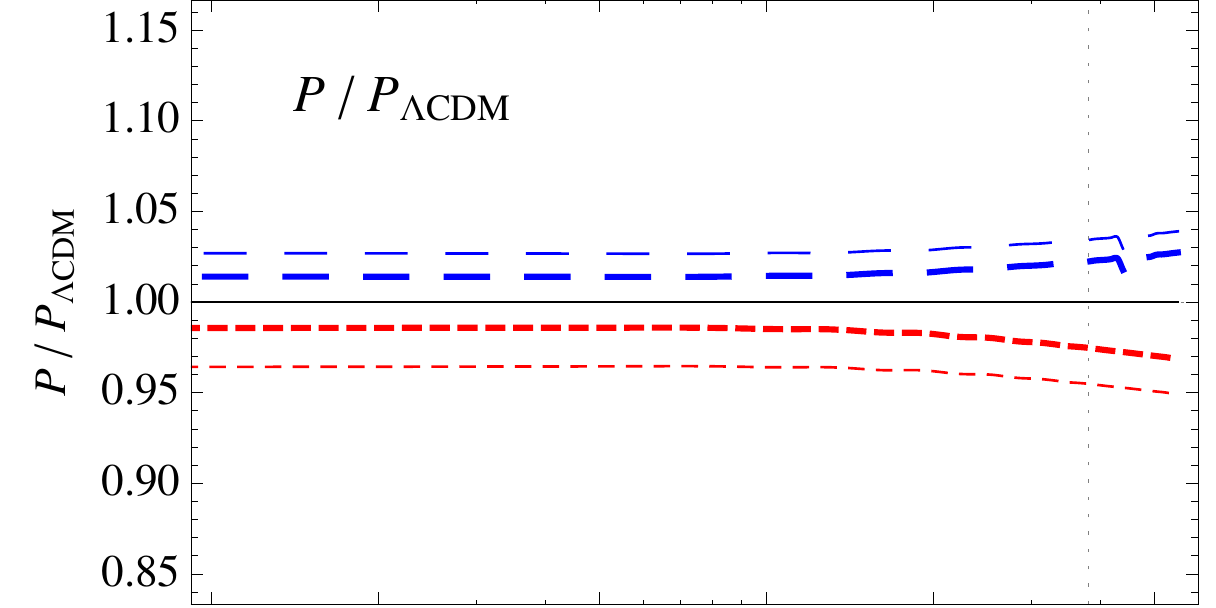}}
{\includegraphics[width=0.49\textwidth]{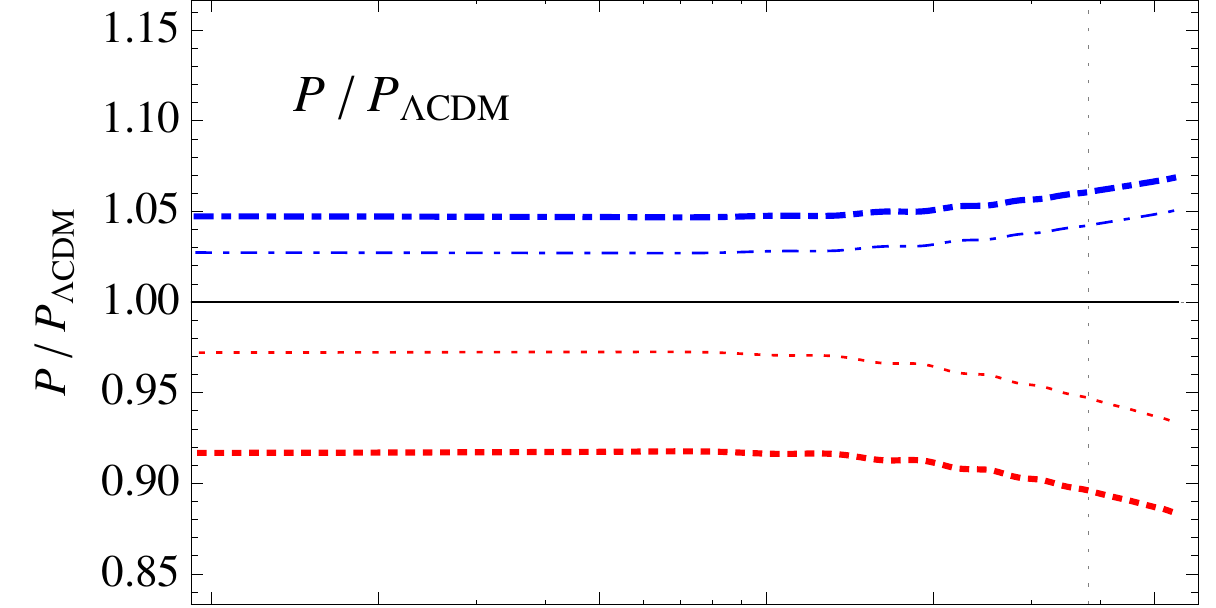}}\\
{\includegraphics[width=0.49\textwidth]{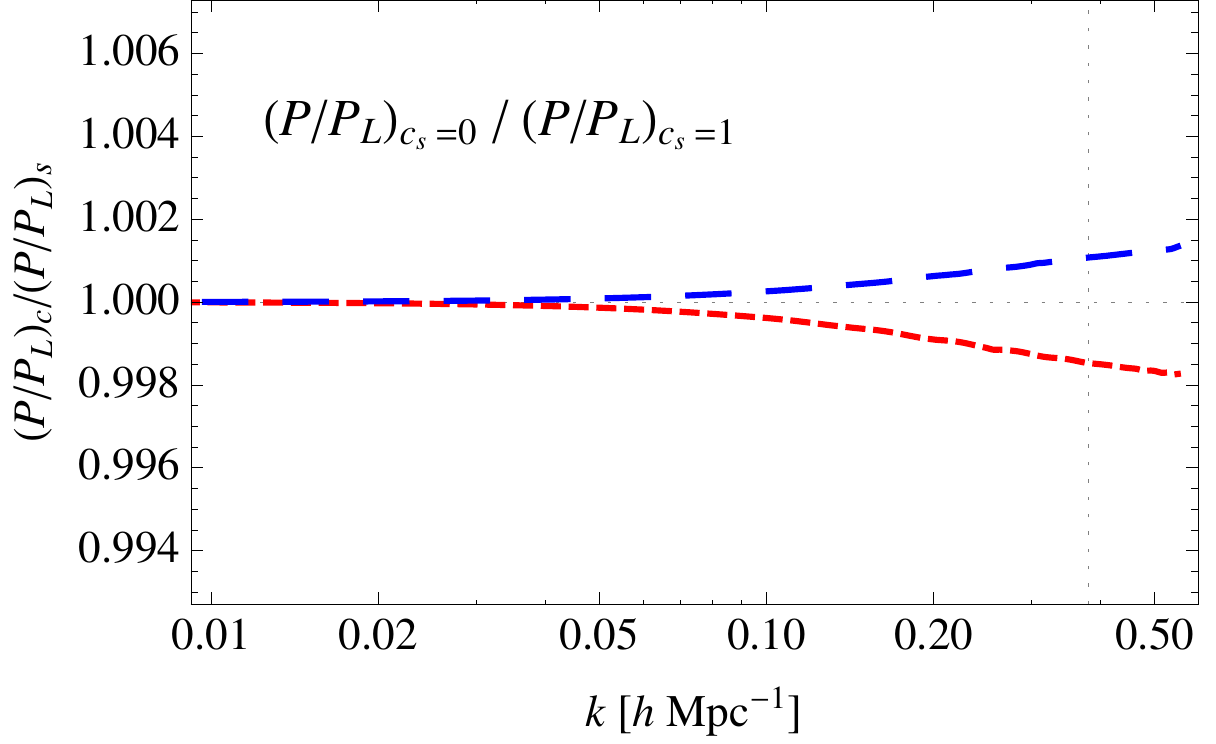}}
{\includegraphics[width=0.49\textwidth]{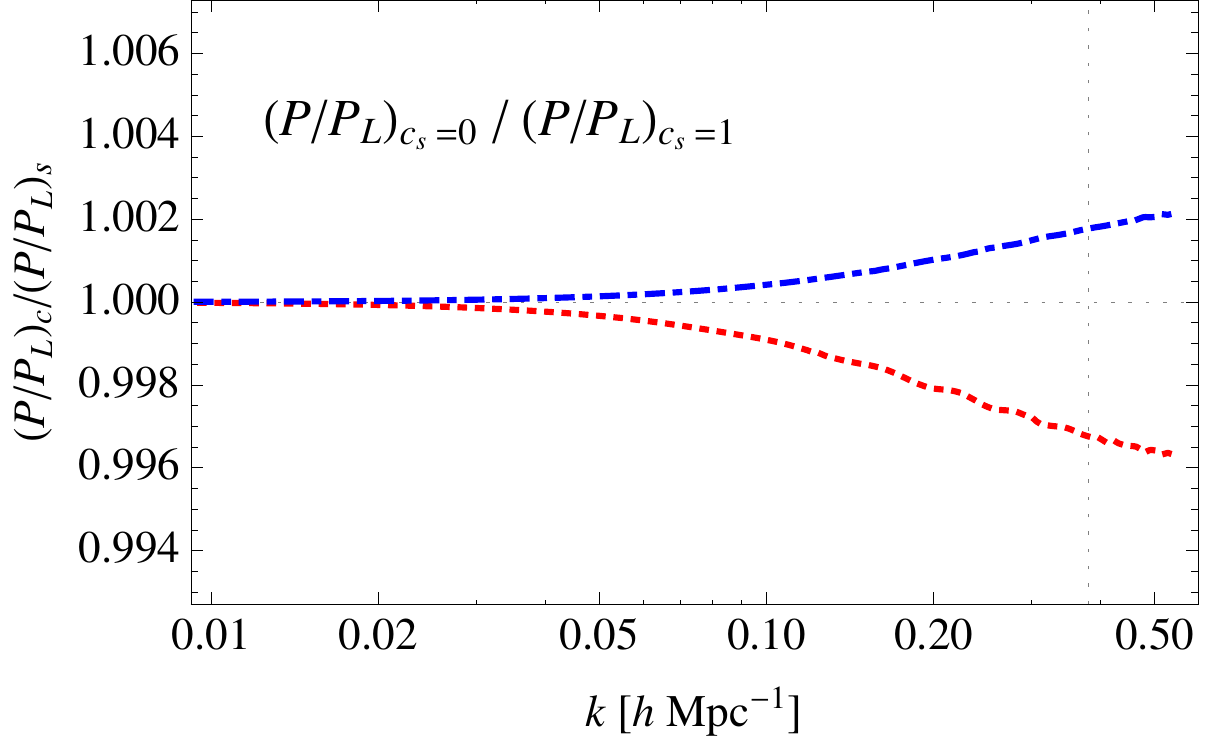}}
\caption{Same as Fig.~\ref{fig:Pz0} but at redshift $z=1$.}
\label{fig:Pz1}
\end{center}
\end{figure}

As previously mentioned, all our predictions assume the same {\em initial} power spectrum, so that the evolved results will differ both in the linear and nonlinear regimes. Fig.~\ref{fig:Pz0},~\ref{fig:Pz0p5} and \ref{fig:Pz1} show our results at redshift $z=0$, $0.5$ and $1$, respectively. Each figure includes, in the upper panels, the total power spectrum as a function of $k$. The middle panels present the ratio of the power spectrum in the quintessence cosmologies to the $\Lambda$CDM case. Finally, in order to highlight the peculiar effect of a vanishing speed of sound on the nonlinear evolution alone, the lower panels show the {\em ratio} of the {\em ratios} of the nonlinear to linear power spectrum of the clustering to the smooth case, for a given value of the equation of state. In each figure, the left panels consider the cases $w=-1.1$ ({\em blue, long-dashed lines}) and $w=-0.9$ ({\em red, short-dashed lines}) while the right panels show the results for $w=-1.2$ ({\em blue, dot-dashed lines}) and $w=-0.8$ ({\em red, dotted lines}).
The dotted vertical line indicates the value of the wavenumber below which the TRG prediction is accurate at 1\% level in the $\Lambda$CDM case. For the three redshift considered such scale is given by $k=0.15\kMpc$, $k=0.18\kMpc$ and $k=0.34\kMpc$, respectively. We determine these values by a direct comparison with numerical simulations, as detailed in Appendix~\ref{app:accuracy}.

We remark from the start that the corrections induced at the nonlinear level by both smooth and clustering quintessence are of the order of the predictions accuracy for $|w+1|\simeq 0.1$ and only slightly larger for $|w+1|\simeq 0.2$. However, it is reasonable to expect the validity of the theoretical predictions to be more extended when ratios between nonlinear power spectra are considered, as it will be the case in the interpretation of our results. In this respect, our results are providing, in the first place, a qualitative picture of the behavior of the nonlinear power spectrum. We anticipate, nonetheless, that improved theoretical methods are possible and will be considered in future works.

Let us focus on the comparison with the $\Lambda$CDM power spectrum, in the linear regime at redshift zero, shown by the middle panels of Fig.~\ref{fig:Pz0}. In the $c_s=1$ cases, shown by the thin curves, the different cosmic expansion history for $w<-1$ with respect to $w>-1$ results in a lower or, respectively, larger growth of dark matter perturbations. The corresponding nonlinear corrections follow the expectations as again, respectively, lower or larger the nonlinear corrections in the $\Lambda$CDM case. 

The behavior of the power spectrum of the {\em total perturbations} is strikingly different. In fact, the opposite effect on the linear growth of clustering versus smooth quintessence corresponds to the results already shown in Fig.~\ref{fig:CD}, where $D(z)$ is given as a function of redshift. The growth of the total perturbations at small redshift is affected by appearance and growth of quintessence perturbations for $w<-1$, while the opposite effect is at place for $w>-1$. However, taking as an example the $w>-1$ case, despite a larger linear growth factor, the nonlinear corrections are significantly smaller with respect to $\Lambda$CDM. Looking at the lower panels, we notice that, relative to the linear power spectrum, the nonlinear corrections are even smaller in the clustering quintessence case, than in the smooth one. The additional corrections induced by the clustering of quintessence are of the order of a few percents in the mildly nonlinear regime for $|w+1|\sim 0.2$ and they are roughly proportional to the departure from $w=-1$.

The peculiar relation between the linear solution and nonlinear corrections can be explained qualitatively taking into account the results for the second-order solution in standard PT for the total density contrast derived by \citet{SefusattiVernizzi2011}. It is possible to show that in the clustering quintessence case some of the terms contributing to such corrections are proportional to the product of the total density growth factor $D$ with the matter growth factor, $D_m$, that is $\d^{(2)}_{\kv}\sim D\,D_m$, while in the standard scenarios one typically has $\d^{(2)}_{\kv}\sim D^2$. This means that the nonlinear evolution of the total fluctuations is in a large part determined by the nonlinear evolution of the matter density alone, leading to a suppression of the nonlinear evolution with respect to the one expected from the value of the growth factor alone in standard scenarios. 

At higher redshift, the effects of quintessence are clearly milder. In particular, at $z=0.5$, Fig.~\ref{fig:Pz0p5}, a vanishing speed of sounds has the effect, for instance assuming $w=-0.8$, of increasing the growth factor by 10\% with respect to the $c_s=1$ case, and reducing the nonlinear correction by less than 1\% at $k\simeq 0.2\kMpc$. Again, the interesting feature is given by the opposite effect on the linear and on the {\em relative} nonlinear growth. At $z=1$, Fig.~\ref{fig:Pz1}, the relevant effect is on the linear growth alone since nonlinear corrections due to quintessence clustering are essentially negligible.

\begin{figure}[t]
\begin{center}
{\includegraphics[width=0.49\textwidth]{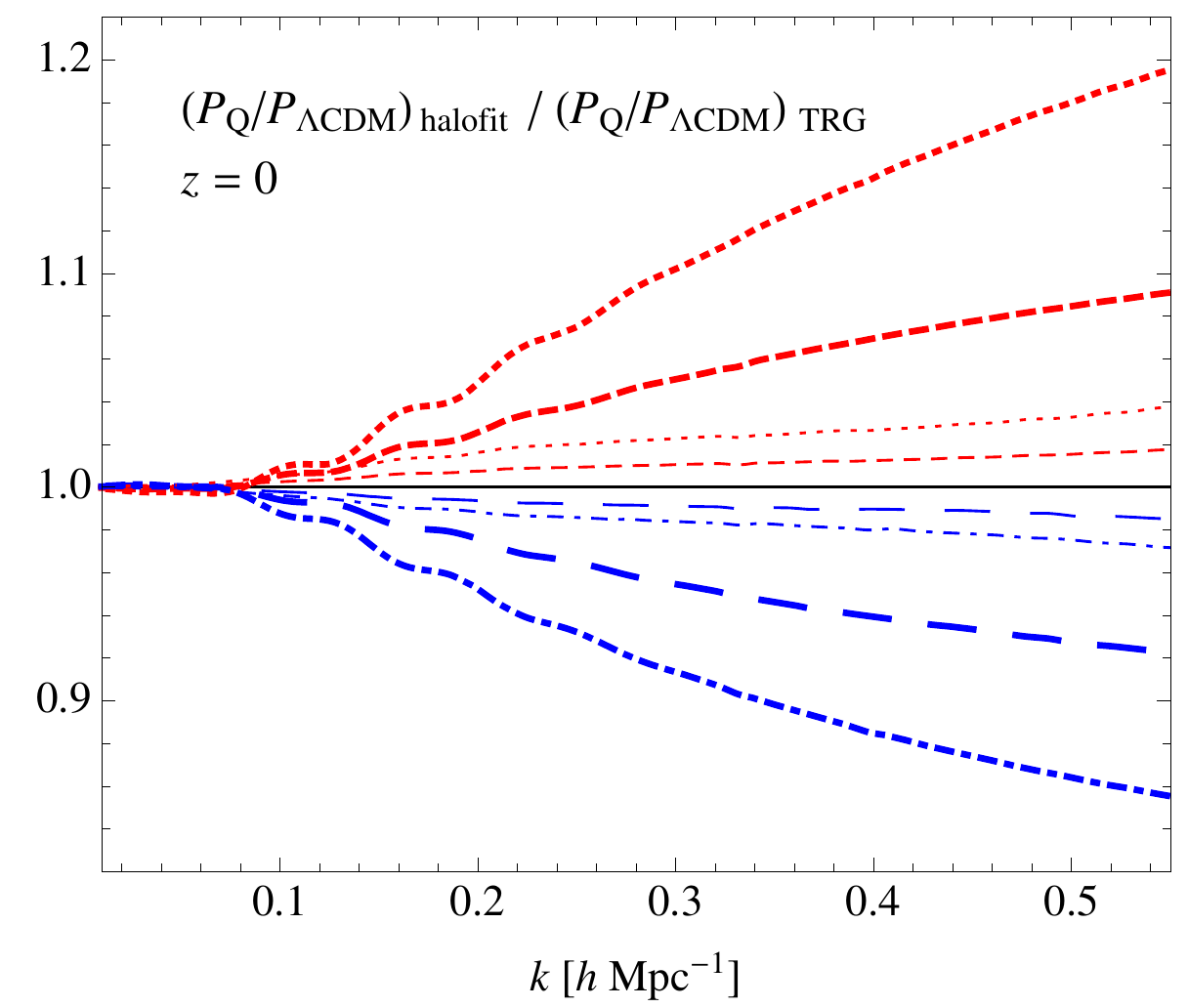}}
{\includegraphics[width=0.49\textwidth]{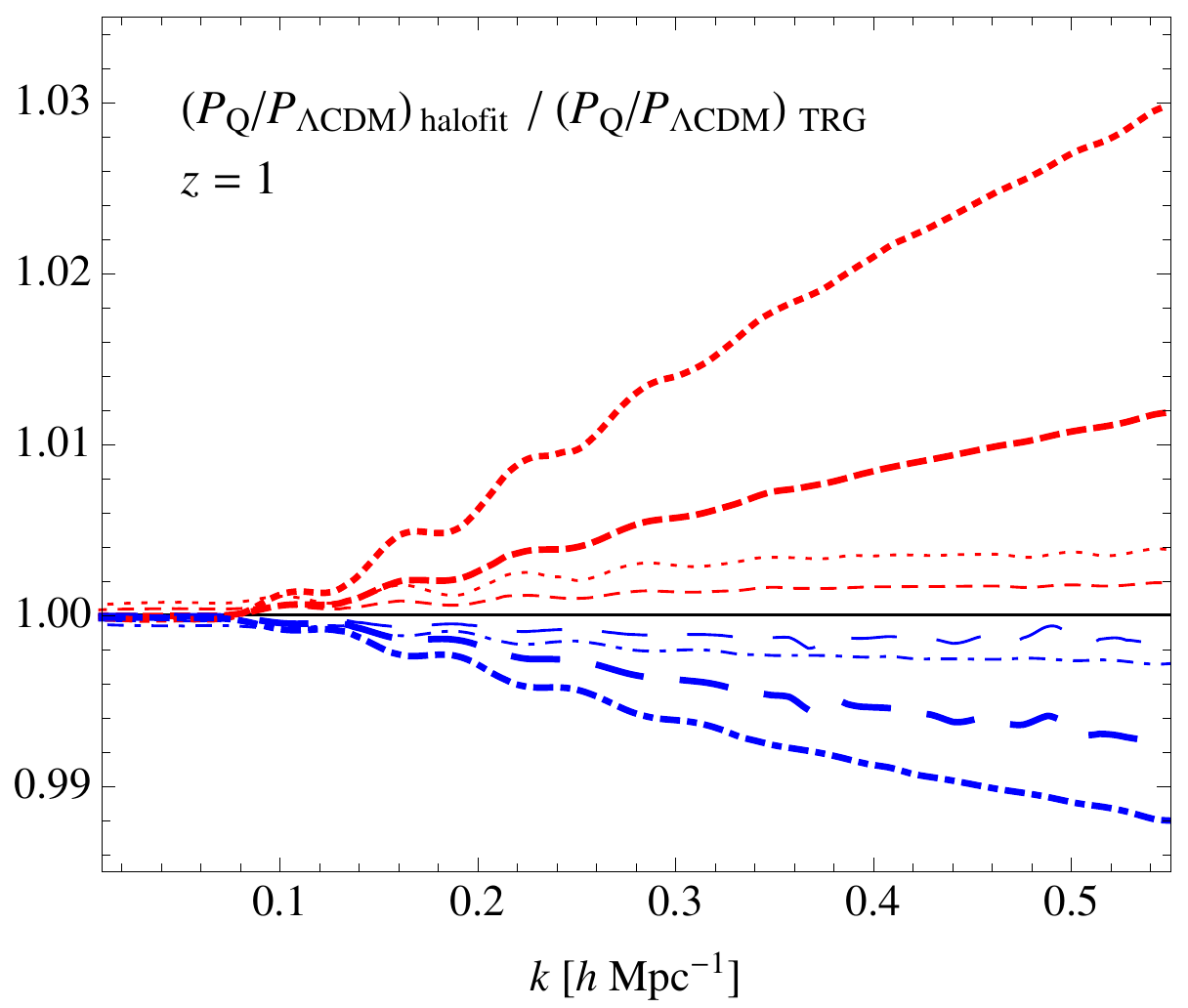}}\\
{\includegraphics[width=0.49\textwidth]{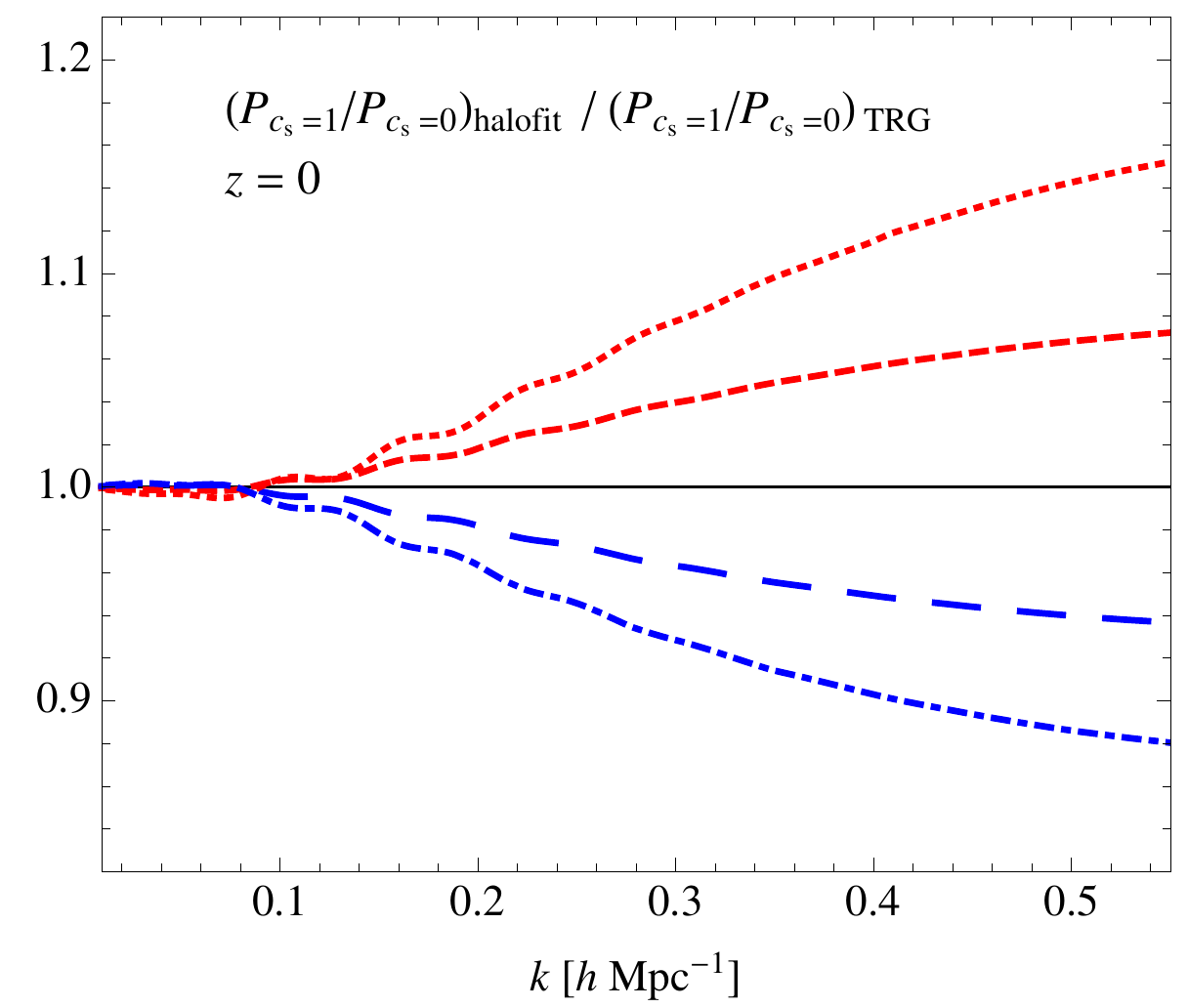}}
{\includegraphics[width=0.49\textwidth]{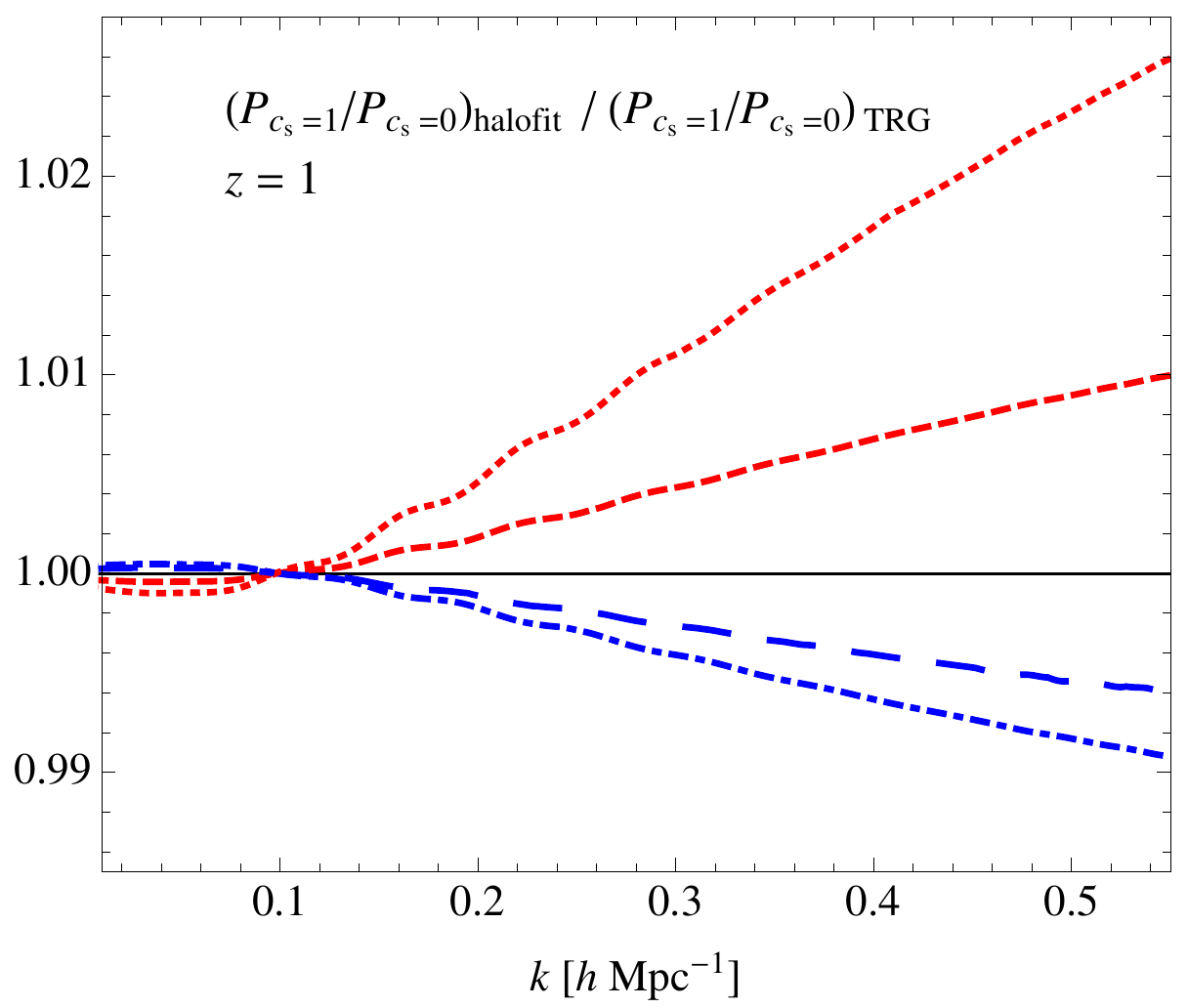}}\\
\caption{{\em Upper panels}: ratio of the ratio between the nonlinear quintessence power spectrum to the $\Lambda$CDM power spectrum predicted by \texttt{halofit} assuming the simple linear prediction, to the ratio of the same quantities computed in the TRG approach at redshift $z=0$ ({\em left panel}) and $z=1$ ({\em right panel}). {\em Lower panels}: ratio of the ratios between the nonlinear quintessence power spectrum for $c_s=1$ to the one for $c_s=0$ predicted by \texttt{halofit} to the ratio of the same quantities computed in the TRG approach. Again, at redshift $z=0$ ({\em left panel}) and $z=1$ ({\em right panel}.}
\label{fig:halofit}
\end{center}
\end{figure}
Because of the peculiar relation between the linear growth factor in clustering quintessence cosmologies and the corresponding nonlinear corrections, simple tools like the \texttt{halofit} fitting function \citep{SmithEtal2003} cannot be directly used for the prediction of the nonlinear power spectrum of the total density perturbations. Still, it is useful to compare such predictions with the Time-RG ones. In fact, \texttt{halofit} provides the expected nonlinear evolution of the power spectrum based on the linear one and, specifically, on the combination of the amplitude of the initial perturbations and their linear growth. To illustrate this point, in the upper panels of Fig.~\ref{fig:halofit} we show a comparison ({\em i.e.} the ratio) between, in turn, the ratio of the nonlinear power spectrum in a quintessence cosmology and the $\Lambda$CDM nonlinear power spectrum computed using \texttt{halofit} and the same quantity computed in the TRG approach. We consider, in particular, the results at redshift $z=0$ ({\em left panel}) and $z=1$ ({\em right panel}). We notice for instance that, due to the larger value of the growth factor at redshift zero, corresponding to a larger value for $\sigma_8$, in the clustering quintessence case, \texttt{halofit} over-predicts (under-predicts) the nonlinear evolution by up to 5\% at $k\simeq 0.2\kMpc$ for $w=-0.8$ ($w=-1.2$). And a much larger discrepancy can be expected at smaller scales. In general, while in the smooth quintessence case the errors are relatively small (below 1\%) on the scales of interest, in the clustering case they can be still of a few percents for $k\simeq 0.4$ even at $z=1$.
It is reasonable to expect that a simple modification to the \texttt{halofit} prescription, perhaps computing the one-halo contribution simply in terms of the {\em matter} linear power spectrum, can extend its validity to the clustering models.  We do not investigate this issue further in this work.

In standard scenarios it is reasonable to expect that to larger values for the linear power spectrum correspond larger nonlinear corrections. As we have seen, such expectations can be disattended, to a certain extent, if quintessence clusters. In the lower panels of Fig.~\ref{fig:halofit} we show instead the ratio between, in turn, the ratio of the nonlinear power spectrum in a quintessence cosmology with $c_s=1$ to the one for $c_s=0$ computed using \texttt{halofit} and the same quantity computed in the TRG approach at redshift $z=0$ ({\em left panel}) and $z=1$ ({\em right panel}). The difference between the two calculations can be taken to a first approximation as the difference expected for the nonlinear corrections for the two values of the sound speed assuming the same linear power spectrum. If indeed the observed density fluctuations result from two comoving fluid as those of the system considered here, the difference in the nonlinear evolution can be as large as ten percent at $z=0$ and still a few percent at $z=1$ for $|w+1|\simeq 0.2$ and approximately proportional to $w+1$.

Before concluding this section, it is interesting to quantify, within the limits of our approximations, the effect that the additional nonlinear evolution induced by the quintessence perturbations has on the position of the acoustic peaks in the total power spectrum. Typically, nonlinear corrections partially smooth out the acoustic features present in the initial power spectrum and induce a shift in the position of the maximum of the correlation function \citep[see, for instance][]{CrocceScoccimarro2008}. We consider here simply the position of the power spectrum maximum at about $k\simeq 0.072\kMpc$ in the ratio of $P(k)$ with respect to a wiggle-less power spectrum in the Time-RG predictions for clustering and smooth quintessence. We find that, at redshift zero, for $w=-0.8$ the position of the peak changes by a $0.5\%$ in the clustering with respect to the smooth case, in fact being closer to the linear theory value, due to the relatively smaller nonlinear growth. For $w=-1.2$, the shift of the maximum between the smooth and clustering quintessence cases is of $0.6\%$ with quintessence perturbations increasing the overall shift with respect to linear theory. At redshift $z=0.5$ the shift is much smaller, below $0.1\%$ for $|w+1|\simeq 0.2$. We remind the reader that these results are obtained assuming the same initial power spectrum and the same value of $\Omega_{m,0}$ for all different models varying only the dark energy equation of state parameter $w$, leading therefore to different values of the linear power spectrum amplitude at redshift zero, \ie different values of $\sigma_8$.

\section{Conclusions}
\label{sec:conclusions}

We studied the nonlinear evolution of the total density power spectrum in cosmologies where the acceleration of the Universe is due to a quintessence field with vanishing speed of sound. As noted in \citet{SefusattiVernizzi2011}, under this assumption, matter and quintessence perturbations are comoving and it is possible to define total density perturbations as a weighted sum of the individual matter and quintessence overdensities. This quantity is particularly relevant since it is directly related to cosmological observables such as cosmic shear or the galaxy power spectrum, while there is no way of distinguishing dark matter from quintessence perturbations as they both only interact gravitationally with baryonic matter.

We compute the nonlinear power spectrum of total matter and quintessence density in the Time-RG approach introduced by \citet{Pietroni2008}. This method is particularly suitable to this problem since the effect of a clustering quintessence field results in a modification of the standard equations of gravitational instability in terms of a single time-dependent factor in the continuity equation, \eqref{continuity_tot}, given by the function $C(z)=1+(1+w)\Omega_Q(z)/\Omega_m(z)$. In the TRG framework, continuity and Euler equations are used to obtain differential equations in time for the density correlators, making the extension to the total density including quintessence with a vanishing speed of sound straightforward. 

Assuming the same initial power spectrum, the difference in the linear growth of the {\em total} density fluctuations between the clustering, $c_s=0$, and the smooth, $c_s=1$, quintessence case is significant at low redshift, being up to 5\% for $|w+1|\simeq 0.1$ at redshift zero. We have shown, however, that the difference between corresponding nonlinear corrections, that is the ratio between nonlinear and linear power spectrum in each case, is relatively small at mildly nonlinear scales, of the order of 1\% at $k=0.15\kMpc$ and redshift zero. This fact is quite interesting because the large difference in the linear regime typically corresponds to larger, by a factor of a few, nonlinear corrections in standard cosmologies with clustering of dark matter alone. The matter component plays a predominant role in the nonlinear evolution of the coupled system and nonlinear corrections to the total power spectrum are in fact very close to those present for the smooth case. As shown by the middle panels of Fig.~\ref{fig:Pz0}, a quintessence cosmology with $w>-1$ and $c_s=0$, leads to a larger growth factor and therefore a larger linear power spectrum with respect to a $\Lambda$CDM cosmology. {\em At the same time}, however, it presents a {\em smaller} power spectrum then $\Lambda$CDM at nonlinear scales due to a suppressed nonlinear evolution (which is in fact very close to the one obtained in the smooth quintessence case). For $c_s=1$, on the other hand, $w>-1$ leads to a smaller linear growth and a correspondingly smaller nonlinear evolution, as usually expected. This peculiar behavior of gravitational instabilities in the presence of dark energy perturbations might have some specific observational consequences that will be the subject of forthcoming work.

During completion of this work, we became aware of a similar work in preparation, \citet{AnselmiBallesterosPietroni2011}, whose contents are not known to us at the time of writing.

\bigskip
{\bf Acknowledgments}\\

We thank Filippo Vernizzi and Rom\'an Scoccimarro for useful discussions and Mart\`in Crocce and Vincent Desjacques for providing the RPT predictions and simulations data for the Appendix. E.S. acknowledges support by the European Commission under the Marie Curie Inter European Fellowship and he is grateful to the Center for Cosmology and Particle Physics of New York University for kind hospitality during the completion of this project.
The work of G.D'A. is supported by a James Arthur Fellowship. We are grateful to the organizers of the PONT 2011 conference, where part of this project has been completed.

\appendix

\section{On the accuracy and validity of the Time-RG approach}
\label{app:accuracy}

\begin{figure}[t]
\begin{center}
{\includegraphics[width=0.49\textwidth]{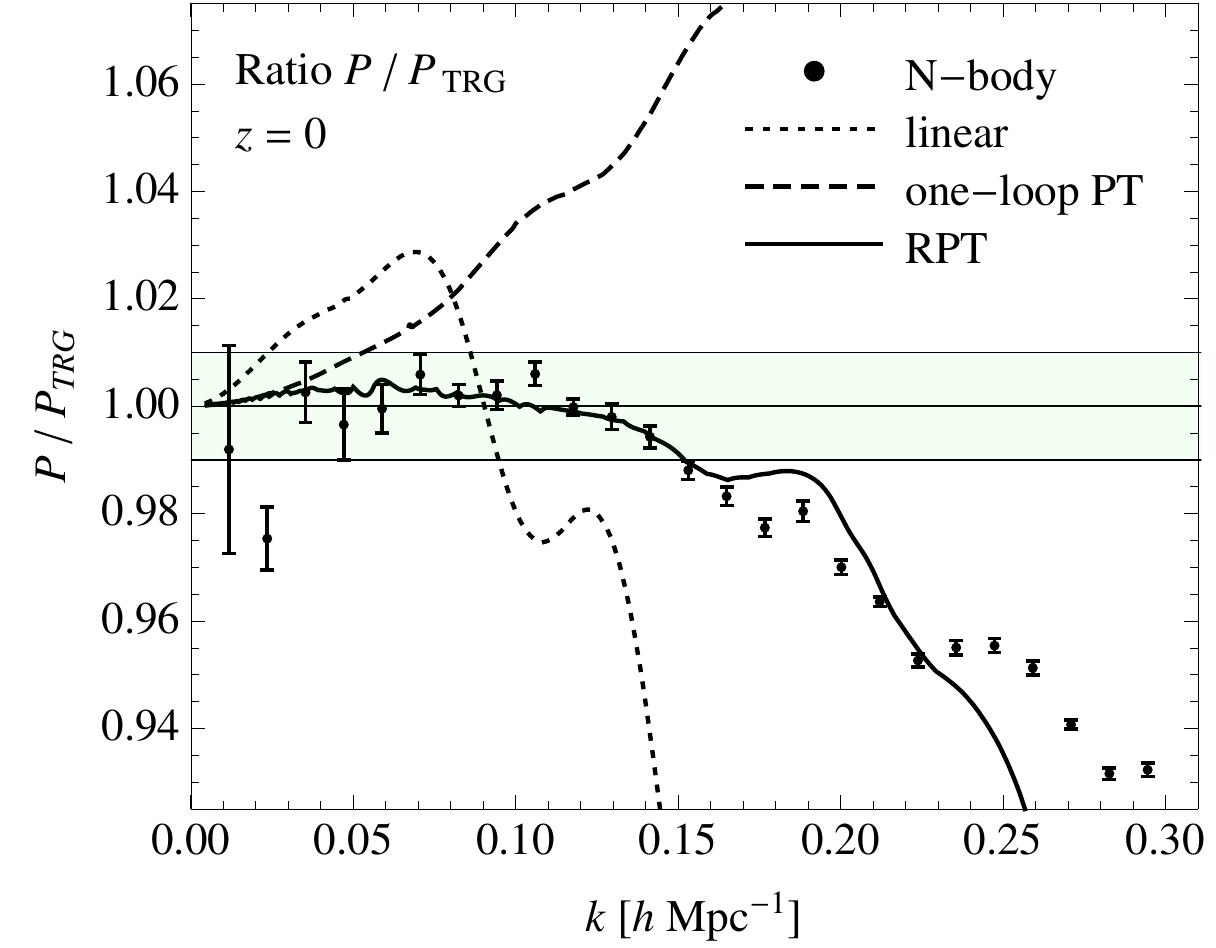}}
{\includegraphics[width=0.49\textwidth]{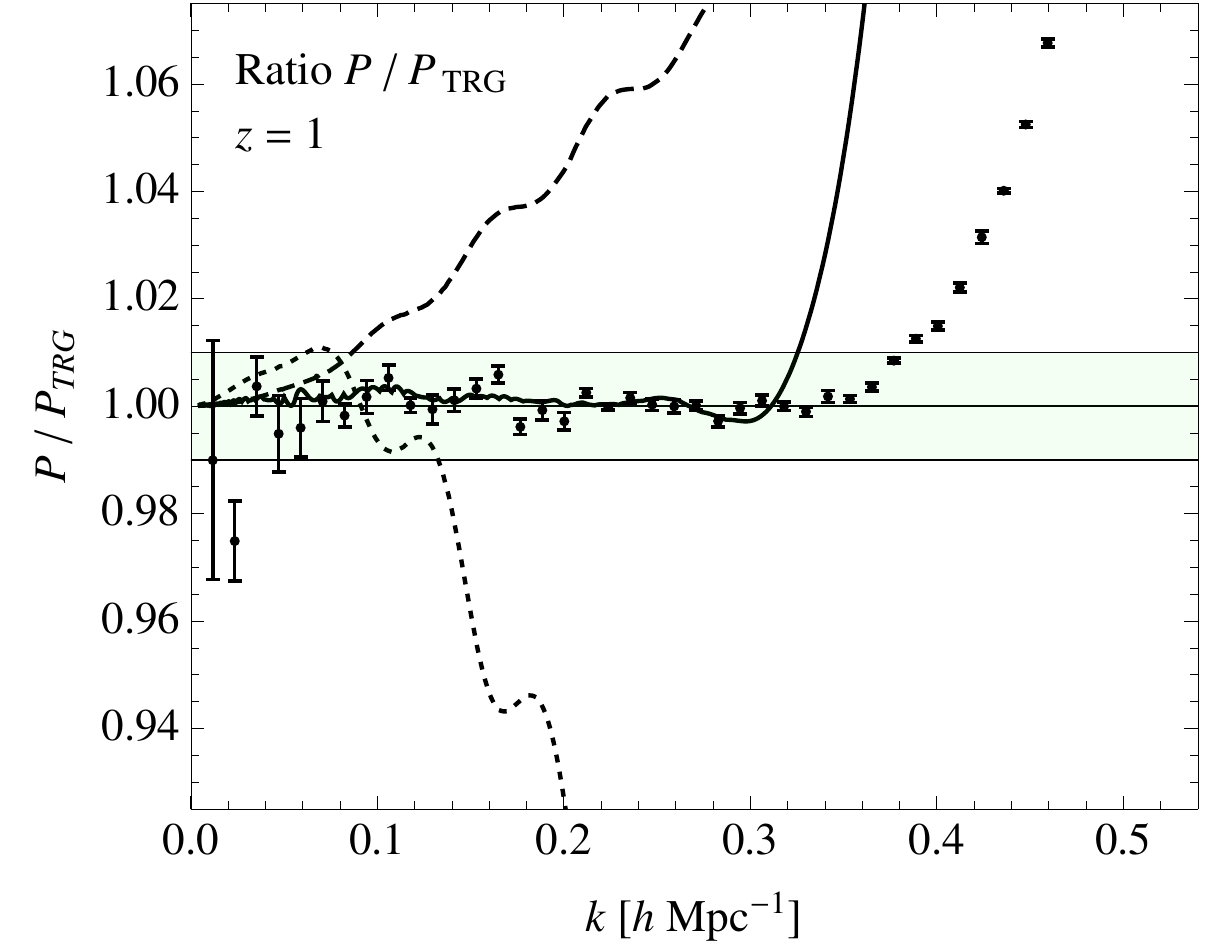}}
\caption{Ratio of the matter power spectrum measured in N-body simulations with a $\Lambda$CDM cosmology to the TRG predictions at redshift $z=0$ ({\em left panel}) and $z=1$ ({\em right panel}). Also shown are the ratio to TRG of the linear theory prediction ({\em dotted curves}), one-loop PT ({\em dashed curves}) and RPT ({\em continuous curves}). Shaded area denotes 1\% accuracy.}
\label{fig:Nbody}
\end{center}
\end{figure}
The accuracy and validity of the TRG predictions for the power spectrum can, so far, only be tested in standard scenarios as no simulations involving a scalar field with vanishing speed of sound are known to the authors at time of writing. In this appendix we simply compare the predictions at redshift $z=0$ and $z=1$ for the $\Lambda$CDM cosmology assumed in our results, with N-body simulations \citep{DesjacquesSeljakIliev2009} as well as with predictions in RPT \citep{CrocceScoccimarro2006A}\footnote{We thank Vincent Desjacques and Martin Crocce for providing us, respectively, with the power spectrum measurements and the RPT predictions.}. 

Figure~\ref{fig:Nbody} shows the ratio of the matter power spectrum measured in N-body simulations with a $\Lambda$CDM cosmology to the TRG predictions at redshift $z=0$ ({\em left panel}) and $z=1$ ({\em right panel}). Also shown are the ratio to TRG of the linear theory prediction ({\em dotted curves}), one-loop PT ({\em dashed curves}) and RPT ({\em continuous curves}). The shaded area denotes 1\% accuracy. We find, in fact, that TRG provides, in this case, predictions accurate to 1\% or better for scales corresponding to $k<0.15\kMpc$, $k<0.19\kMpc$ and $k< 0.38\kMpc$ respectively at $z=0$, $0.5$ and $z=1$. These results are somehow at odds with the results of \citet{CarlsonWhitePadmanabhan2009} where lowest wavenumbers corresponding to a 1\% accuracy are reported to be $0.04$ and $0.09\kMpc$ respectively for $z=0$ and $1$. In addition, as shown in the figure, we find that the RPT approach, in this case evaluated at the two-loop approximation, provides a prediction better than 1\% up to $0.24\kMpc$ at redshift zero and up to $0.32\kMpc$ at $z=1$, in agreement with \citet{CrocceScoccimarro2008}. An extensive investigation of the discrepancies between different tests of resummation methods in PT is clearly beyond the scope of this work.

\bigskip

\bibliography{Bibliography}

\begin{thebibliography}{41}
\newcommand{\enquote}[1]{``#1''}
\providecommand{\natexlab}[1]{#1}
\providecommand{\url}[1]{\texttt{#1}}
\providecommand{\urlprefix}{URL }
\providecommand{\eprint}[2][]{\url{#2}}

\bibitem[{Anselmi \textit{et~al.}(2011)Anselmi, Ballesteros, \&
  Pietroni}]{AnselmiBallesterosPietroni2011}
Anselmi, S., Ballesteros, G., \& Pietroni, M., 2011.
\newblock \textit{{\em in preparation}}

\bibitem[{Arkani-Hamed \textit{et~al.}(2004)Arkani-Hamed, {Cheng}, {Luty}, \&
  {Mukohyama}}]{ArkaniHamedEtal2004A}
Arkani-Hamed, N., {Cheng}, H.~S., {Luty}, M.~A., \& {Mukohyama}, S., 2004.
\newblock \enquote{{Ghost Condensation and a Consistent IR Modification of
  Gravity}.}
\newblock \textit{Journal of High Energy Physics}, \textbf{5}, 74.
\newblock \eprint{arXiv:hep-th/0312099}

\bibitem[{{Armendariz-Picon} \textit{et~al.}(2000){Armendariz-Picon},
  {Mukhanov}, \& {Steinhardt}}]{ArmendarizPiconMukhanovSteinhardt2000}
{Armendariz-Picon}, C., {Mukhanov}, V., \& {Steinhardt}, P.~J., 2000.
\newblock \enquote{{Dynamical Solution to the Problem of a Small Cosmological
  Constant and Late-Time Cosmic Acceleration}.}
\newblock \textit{Physical Review Letters}, \textbf{85}, 4438.
\newblock \eprint{arXiv:astro-ph/0004134}

\bibitem[{{Bartolo} \textit{et~al.}(2010){Bartolo}, {Beltr{\'a}n Almeida},
  {Matarrese}, {Pietroni}, \& {Riotto}}]{BartoloEtal2010}
{Bartolo}, N., {Beltr{\'a}n Almeida}, J.~P., {Matarrese}, S., {Pietroni}, M.,
  \& {Riotto}, A., 2010.
\newblock \enquote{{Signatures of primordial non-Gaussianities in the matter
  power-spectrum and bispectrum: the time-RG approach}.}
\newblock \textit{\jcap}, \textbf{3}, 11.
\newblock \eprint{arXiv: 0912.4276 [astro-ph.CO]}

\bibitem[{Bean \& Dor{\'e}(2004)}]{BeanDore2004}
Bean, R. \& Dor{\'e}, O., 2004.
\newblock \enquote{Probing dark energy perturbations: The dark energy equation
  of state and speed of sound as measured by wmap.}
\newblock \textit{\prd}, \textbf{69}, 8, 083503.
\newblock \eprint{arXiv:astro-ph/0307100}

\bibitem[{Bernardeau \textit{et~al.}(2002)Bernardeau, Colombi, Gazta{\~n}aga,
  \& Scoccimarro}]{BernardeauEtal2002}
Bernardeau, F., Colombi, S., Gazta{\~n}aga, E., \& Scoccimarro, R., 2002.
\newblock \enquote{Large-scale structure of the universe and cosmological
  perturbation theory.}
\newblock \textit{\physrep}, \textbf{367}, 1.
\newblock \eprint{arXiv: astro-ph/0112551}

\bibitem[{Bernardeau \textit{et~al.}(2008)Bernardeau, Crocce, \&
  Scoccimarro}]{BernardeauCrocceScoccimarro2008}
Bernardeau, F., Crocce, M., \& Scoccimarro, R., 2008.
\newblock \enquote{Multipoint propagators in cosmological gravitational
  instability.}
\newblock \textit{\prd}, \textbf{78}, 10, 103521.
\newblock \eprint{arXiv: 0806.2334}

\bibitem[{Bernardeau \textit{et~al.}(2010)Bernardeau, Crocce, \&
  Sefusatti}]{BernardeauCrocceSefusatti2010}
Bernardeau, F., Crocce, M., \& Sefusatti, E., 2010.
\newblock \enquote{Multipoint propagators for non-gaussian initial conditions.}
\newblock \textit{\prd}, \textbf{82}, 8, 083507.
\newblock \eprint{arXiv: 1006.4656 [astro-ph.CO]}

\bibitem[{Carlson \textit{et~al.}(2009)Carlson, White, \&
  Padmanabhan}]{CarlsonWhitePadmanabhan2009}
Carlson, J., White, M., \& Padmanabhan, N., 2009.
\newblock \enquote{Critical look at cosmological perturbation theory
  techniques.}
\newblock \textit{\prd}, \textbf{80}, 4, 043531.
\newblock \eprint{arXiv: 0905.0479 [astro-ph.CO]}

\bibitem[{Corasaniti \textit{et~al.}(2005)Corasaniti, Giannantonio, \&
  Melchiorri}]{CorasanitiGiannantonioMelchiorri2005}
Corasaniti, P., Giannantonio, T., \& Melchiorri, A., 2005.
\newblock \enquote{{Constraining dark energy with cross-correlated CMB and
  large scale structure data}.}
\newblock \textit{\prd}, \textbf{71}, 12, 123521.
\newblock \eprint{arXiv:astro-ph/0504115}

\bibitem[{Creminelli \textit{et~al.}(2010)Creminelli, D'Amico, Nore{\~n}a,
  Senatore, \& Vernizzi}]{CreminelliEtal2010}
Creminelli, P., D'Amico, G., Nore{\~n}a, J., Senatore, L., \& Vernizzi, F.,
  2010.
\newblock \enquote{Spherical collapse in quintessence models with zero speed of
  sound.}
\newblock \textit{\jcap}, \textbf{3}, 27.
\newblock \eprint{arXiv: 0911.2701 [astro-ph.CO]}

\bibitem[{Creminelli \textit{et~al.}(2009)Creminelli, D'Amico, Nore{\~n}a, \&
  Vernizzi}]{CreminelliEtal2009}
Creminelli, P., D'Amico, G., Nore{\~n}a, J., \& Vernizzi, F., 2009.
\newblock \enquote{The effective theory of quintessence: the {$w<-1$} side
  unveiled.}
\newblock \textit{\jcap}, \textbf{2}, 18.
\newblock \eprint{arXiv: 0811.0827}

\bibitem[{Crocce \& Scoccimarro(2006{\natexlab{a}})}]{CrocceScoccimarro2006B}
Crocce, M. \& Scoccimarro, R., 2006{\natexlab{a}}.
\newblock \enquote{Memory of initial conditions in gravitational clustering.}
\newblock \textit{\prd}, \textbf{73}, 6, 063520.
\newblock \eprint{arXiv: astro-ph/0509419}

\bibitem[{Crocce \& Scoccimarro(2006{\natexlab{b}})}]{CrocceScoccimarro2006A}
Crocce, M. \& Scoccimarro, R., 2006{\natexlab{b}}.
\newblock \enquote{Renormalized cosmological perturbation theory.}
\newblock \textit{\prd}, \textbf{73}, 6, 063519.
\newblock \eprint{arXiv: astro-ph/0509418}

\bibitem[{Crocce \& Scoccimarro(2008)}]{CrocceScoccimarro2008}
Crocce, M. \& Scoccimarro, R., 2008.
\newblock \enquote{Nonlinear evolution of baryon acoustic oscillations.}
\newblock \textit{\prd}, \textbf{77}, 2, 023533.
\newblock \eprint{arXiv: 0704.2783}

\bibitem[{DeDeo \textit{et~al.}(2003)DeDeo, Caldwell, \&
  Steinhardt}]{DeDeoCaldwellSteinhardt2003}
DeDeo, S., Caldwell, R.~R., \& Steinhardt, P.~J., 2003.
\newblock \enquote{Effects of the sound speed of quintessence on the microwave
  background and large scale structure.}
\newblock \textit{\prd}, \textbf{67}, 10, 103509.
\newblock \eprint{arXiv:astro-ph/0301284}

\bibitem[{Desjacques \textit{et~al.}(2009)Desjacques, Seljak, \&
  Iliev}]{DesjacquesSeljakIliev2009}
Desjacques, V., Seljak, U., \& Iliev, I.~T., 2009.
\newblock \enquote{Scale-dependent bias induced by local non-gaussianity: a
  comparison to n-body simulations.}
\newblock \textit{\mnras}, 631.
\newblock \eprint{arXiv: 0811.2748}

\bibitem[{Ferreira \& Joyce(1997)}]{FerreiraJoyce1997}
Ferreira, P.~G. \& Joyce, M., 1997.
\newblock \enquote{{Structure Formation with a Self-Tuning Scalar Field}.}
\newblock \textit{Physical Review Letters}, \textbf{79}, 4740.
\newblock \eprint{arXiv:astro-ph/9707286}

\bibitem[{Hannestad(2005)}]{Hannestad2005}
Hannestad, S., 2005.
\newblock \enquote{Constraints on the sound speed of dark energy.}
\newblock \textit{\prd}, \textbf{71}, 10, 103519.
\newblock \eprint{arXiv:astro-ph/0504017}

\bibitem[{Hu \& Scranton(2004)}]{HuScranton2004}
Hu, W. \& Scranton, R., 2004.
\newblock \enquote{Measuring dark energy clustering with cmb-galaxy
  correlations.}
\newblock \textit{\prd}, \textbf{70}, 12, 123002.
\newblock \eprint{arXiv:astro-ph/0408456}

\bibitem[{Lesgourgues \textit{et~al.}(2009)Lesgourgues, Matarrese, Pietroni, \&
  Riotto}]{LesgouguesEtal2009}
Lesgourgues, J., Matarrese, S., Pietroni, M., \& Riotto, A., 2009.
\newblock \enquote{Non-linear power spectrum including massive neutrinos: the
  time-rg flow approach.}
\newblock \textit{\jcap}, \textbf{6}, 17.
\newblock \eprint{arXiv: 0901.4550 [astro-ph.CO]}

\bibitem[{Lim \textit{et~al.}(2010)Lim, {Sawicki}, \&
  {Vikman}}]{LimSawickiVikman2010}
Lim, E.~A., {Sawicki}, I., \& {Vikman}, A., 2010.
\newblock \enquote{Dust of dark energy.}
\newblock \textit{\jcap}, \textbf{5}, 12.
\newblock \eprint{arXiv: 1003.5751 [astro-ph.CO]}

\bibitem[{Matarrese \& Pietroni(2007)}]{MatarresePietroni2007}
Matarrese, S. \& Pietroni, M., 2007.
\newblock \enquote{Resumming cosmic perturbations.}
\newblock \textit{Journal of Cosmology and Astro-Particle Physics}, \textbf{6},
  26.
\newblock \eprint{arXiv:astro-ph/0703563}

\bibitem[{Matarrese \& Pietroni(2008)}]{MatarresePietroni2008}
Matarrese, S. \& Pietroni, M., 2008.
\newblock \enquote{{Baryonic Acoustic Oscillations via the Renormalization
  Group}.}
\newblock \textit{Modern Physics Letters A}, \textbf{23}, 25.
\newblock \eprint{arXiv:astro-ph/0702653}

\bibitem[{Matsubara(2008{\natexlab{a}})}]{Matsubara2008Berr}
Matsubara, T., 2008{\natexlab{a}}.
\newblock \enquote{{Erratum: Nonlinear perturbation theory with halo bias and
  redshift-space distortions via the Lagrangian picture [Phys. Rev. D 78,
  083519 (2008)]}.}
\newblock \textit{\prd}, \textbf{78}, 10, 109901

\bibitem[{Matsubara(2008{\natexlab{b}})}]{Matsubara2008B}
Matsubara, T., 2008{\natexlab{b}}.
\newblock \enquote{{Nonlinear perturbation theory with halo bias and
  redshift-space distortions via the Lagrangian picture}.}
\newblock \textit{\prd}, \textbf{78}, 8, 083519.
\newblock \eprint{arXiv: 0807.1733}

\bibitem[{Matsubara(2008{\natexlab{c}})}]{Matsubara2008A}
Matsubara, T., 2008{\natexlab{c}}.
\newblock \enquote{{Resumming cosmological perturbations via the Lagrangian
  picture: One-loop results in real space and in redshift space}.}
\newblock \textit{\prd}, \textbf{77}, 6, 063530.
\newblock \eprint{arXiv: 0711.2521}

\bibitem[{Matsubara(2011)}]{Matsubara2011}
Matsubara, T., 2011.
\newblock \enquote{Nonlinear perturbation theory integrated with nonlocal bias,
  redshift-space distortions, and primordial non-gaussianity.}
\newblock \textit{ArXiv e-prints}.
\newblock \eprint{arXiv: 1102.4619 [astro-ph.CO]}

\bibitem[{Pietroni(2008)}]{Pietroni2008}
Pietroni, M., 2008.
\newblock \enquote{Flowing with time: a new approach to non-linear cosmological
  perturbations.}
\newblock \textit{Journal of Cosmology and Astro-Particle Physics},
  \textbf{10}, 36.
\newblock \eprint{arXiv: 0806.0971}

\bibitem[{Sapone \& Kunz(2009)}]{SaponeKunz2009}
Sapone, D. \& Kunz, M., 2009.
\newblock \enquote{Fingerprinting dark energy.}
\newblock \textit{\prd}, \textbf{80}, 8, 083519.
\newblock \eprint{arXiv: 0909.0007 [astro-ph.CO]}

\bibitem[{Sapone \textit{et~al.}(2010)Sapone, Kunz, \&
  Amendola}]{SaponeKunzAmendola2010}
Sapone, D., Kunz, M., \& Amendola, L., 2010.
\newblock \enquote{Fingerprinting dark energy. ii. weak lensing and galaxy
  clustering tests.}
\newblock \textit{\prd}, \textbf{82}, 10, 103535.
\newblock \eprint{arXiv: 1007.2188 [astro-ph.CO]}

\bibitem[{Scoccimarro(1998)}]{Scoccimarro1998}
Scoccimarro, R., 1998.
\newblock \enquote{Transients from initial conditions: a perturbative
  analysis.}
\newblock \textit{\mnras}, \textbf{299}, 1097.
\newblock \eprint{arXiv:astro-ph/9711187}

\bibitem[{Sefusatti \& Vernizzi(2011)}]{SefusattiVernizzi2011}
Sefusatti, E. \& Vernizzi, F., 2011.
\newblock \enquote{Cosmological structure formation with clustering
  quintessence.}
\newblock \textit{\jcap}, \textbf{3}, 47.
\newblock \eprint{arXiv: 1101.1026 [astro-ph.CO]}

\bibitem[{Smith \textit{et~al.}(2003)Smith, {Peacock}, {Jenkins}, White,
  {Frenk}, {Pearce}, {Thomas}, {Efstathiou}, \& {Couchman}}]{SmithEtal2003}
Smith, R.~E., {Peacock}, J.~A., {Jenkins}, A., White, S. D.~M., {Frenk}, C.~S.,
  {Pearce}, F.~R., {Thomas}, P.~A., {Efstathiou}, G., \& {Couchman}, H.~M.~P.,
  2003.
\newblock \enquote{{Stable clustering, the halo model and non-linear
  cosmological power spectra}.}
\newblock \textit{\mnras}, \textbf{341}, 1311.
\newblock \eprint{arXiv:astro-ph/0207664}

\bibitem[{Takada(2006)}]{Takada2006}
Takada, M., 2006.
\newblock \enquote{Can a galaxy redshift survey measure dark energy
  clustering?}
\newblock \textit{\prd}, \textbf{74}, 4, 043505.
\newblock \eprint{arXiv:astro-ph/0606533}

\bibitem[{Taruya \& Hiramatsu(2008)}]{TaruyaHiramatsu2008}
Taruya, A. \& Hiramatsu, T., 2008.
\newblock \enquote{A closure theory for nonlinear evolution of cosmological
  power spectra.}
\newblock \textit{\apj}, \textbf{674}, 617.
\newblock \eprint{arXiv: 0708.1367}

\bibitem[{{Torres-Rodr{\'{\i}}guez} \&
  {Cress}(2007)}]{Torres-RodriguezCress2007}
{Torres-Rodr{\'{\i}}guez}, A. \& {Cress}, C.~M., 2007.
\newblock \enquote{Constraining the nature of dark energy using the square
  kilometer array telescope.}
\newblock \textit{\mnras}, \textbf{376}, 1831.
\newblock \eprint{arXiv:astro-ph/0702113}

\bibitem[{Valageas(2007)}]{Valageas2007}
Valageas, P., 2007.
\newblock \enquote{{Large-N expansions applied to gravitational clustering}.}
\newblock \textit{\aap}, \textbf{465}, 725.
\newblock \eprint{arXiv:astro-ph/0611849}

\bibitem[{Wang \& Steinhardt(1998)}]{WangSteinhardt1998}
Wang, L. \& Steinhardt, P.~J., 1998.
\newblock \enquote{Cluster abundance constraints for cosmological models with a
  time-varying, spatially inhomogeneous energy component with negative
  pressure.}
\newblock \textit{\apj}, \textbf{508}, 483.
\newblock \eprint{arXiv:astro-ph/9804015}

\bibitem[{Weller \& {Lewis}(2003)}]{WellerLewis2003}
Weller, J. \& {Lewis}, A.~M., 2003.
\newblock \enquote{Large-scale cosmic microwave background anisotropies and
  dark energy.}
\newblock \textit{\mnras}, \textbf{346}, 987.
\newblock \eprint{arXiv:astro-ph/0307104}

\bibitem[{Zlatev \textit{et~al.}(1999)Zlatev, Wang, \&
  Steinhardt}]{ZlatevWangSteinhardt1999}
Zlatev, I., Wang, L., \& Steinhardt, P.~J., 1999.
\newblock \enquote{Quintessence, cosmic coincidence, and the cosmological
  constant.}
\newblock \textit{Physical Review Letters}, \textbf{82}, 896.
\newblock \eprint{arXiv:astro-ph/9807002}

\end{thebibliography}

\end{document}